\documentclass[acmsmall]{acmart}

\setcopyright{cc}
\setcctype{by}
\acmDOI{10.1145/3832271}
\acmYear{2026}
\acmJournal{PACMSE}
\acmVolume{3}
\acmNumber{ISSTA}
\acmArticle{ISSTA180}
\acmMonth{10}
\acmSubmissionID{issta26main-p2094-p}
\received{2026-01-30}
\received[accepted]{2026-04-16}

\usepackage[linesnumbered,ruled,vlined]{algorithm2e}

\usepackage{textcomp}
\usepackage{caption}
\usepackage{subcaption}
\usepackage{multirow}
\usepackage{graphicx}
\usepackage{booktabs}
\usepackage{enumitem}
\usepackage{listings}
\usepackage{multicol} 
\usepackage{wrapfig}
\usepackage{amsmath}

\usepackage[table]{xcolor}

\definecolor{codegreen}{rgb}{0,0.6,0}
\definecolor{codegray}{rgb}{0.5,0.5,0.5}
\definecolor{codepurple}{rgb}{0.58,0,0.82}
\definecolor{backcolour}{rgb}{0.95,0.95,0.92}

\lstdefinestyle{mystyle}{
  backgroundcolor=\color{backcolour},   
  commentstyle=\color{codegreen},
  keywordstyle=\color{magenta},
  numberstyle=\tiny\color{codegray},
  stringstyle=\color{codepurple},
  basicstyle=\ttfamily\footnotesize,
  breakatwhitespace=false,         
  breaklines=true,                 
  captionpos=b,                    
  keepspaces=true,                 
  numbers=left,                    
  numbersep=5pt,                  
  showspaces=false,                
  showstringspaces=false,
  showtabs=false,                  
  tabsize=2,
  float=!htpb,
  xleftmargin=10pt,
  aboveskip=0pt,               % Remove space above listings
  belowskip=0pt,
}

\begin{document}

\newcommand{\todo}[1]{\textbf{\color{red}{Ravishka: #1} }}

\newcommand{\zoe}[1]{\textcolor{blue}{Zoe: [#1]}}

\newcommand{\wei}[1]{\textcolor{blue}{Wei: [#1]}}

\newcommand{\ww}[1]{\textcolor{blue}{Weihang: [#1]}}

\newcommand{\td}[1]{\textcolor{blue}{ToDo: [#1]}}

\newcommand{\zijie}[1]{{\color{purple}[Zijie: #1]}}

\newcommand{\zeqing}[1]{{\textcolor{blue}[Zeqing: #1]}}

\newcommand{\jiajun}[1]{{\color{cyan}[Jiang: #1]}}

\newcommand{\CR}[1]{\textcolor{blue}{\textbf{#1}}}

\newcommand{\distance}{4pt}
\setlength{\textfloatsep}{\distance}

\newcommand\mycommfont[1]{\small\ttfamily\textcolor{violet}{#1}}
\SetCommentSty{mycommfont}

\lstdefinestyle{Cpp}{ % Define a style for your code snippet, multiple definitions can be made if, for example, you wish to insert multiple code snippets using different programming languages into one document
	%    backgroundcolor=\color{highlight}, % Set the background color for the snippet - useful for highlighting
	language=C++,
	basicstyle=\scriptsize\ttfamily, % The default font size and style of the code
	breakatwhitespace=false, % If true, only allows line breaks at white space
	breaklines=true, % Automatic line breaking (prevents code from protruding outside the box)
	captionpos=b, % Sets the caption position: b for bottom; t for top
	commentstyle=\color[rgb]{0.0, 0.5, 0.69},%\color[rgb]{0,0.6,0}, % Style of comments within the code - dark green courier font
	deletekeywords={}, % If you want to delete any keywords from the current language separate them by commas
	%escapeinside={\%}, % This allows you to escape to LaTeX using the character in the bracket
	escapeinside={<@}{@>},
	firstnumber=1, % Line numbers begin at line 1
	frame=lines, % Frame around the code box, value can be: none, leftline, topline, bottomline, lines, single, shadowbox
	frameround=tttt, % Rounds the corners of the frame for the top left, top right, bottom left and bottom right positions
	keywordstyle={[1]\color{blue!90!black}},
	keywordstyle={[3]\color{red!80!orange}},
	morekeywords={String,int}, % Add any functions no included by default here separated by commas
	numbers=left, % Location of line numbers, can take the values of: none, left, right
	numbersep=-8pt, % Distance of line numbers from the code box
	numberstyle=\tiny\color[rgb]{0.1,0.1,0.1}, % Style used for line numbers
	rulecolor=\color{black}, % Frame border color
	showstringspaces=false, % Don't put marks in string spaces
	showtabs=false, % Display tabs in the code as lines
	stepnumber=1, % The step distance between line numbers, i.e. how often will lines be numbered
	stringstyle=\color[rgb]{0.58,0,0.82},
	tabsize=2, % Number of spaces per tab in the code
	backgroundcolor=\color{white}
}

\title{Characterizing Real-World Bugs in Tile Programs for Automated Bug Detection}  

\author{Ravishka Rathnasuriya}
\orcid{0009-0005-6129-2865}
\authornote{Ravishka Rathnasuriya and Zihe Song contributed equally to this research.}
\email{ravishka.rathnasuriya@utdallas.edu}
\affiliation{%
  \institution{University of Texas at Dallas}
  \city{Richardson}
  % \state{Texas}
  \country{USA}
}

\author{Zihe Song}
\orcid{0009-0009-6651-1560}
\authornotemark[1]
\email{zihe.song@utdallas.edu}
\affiliation{%
  \institution{University of Texas at Dallas}
  \city{Richardson}
  % \state{Texas}
  \country{USA}
}

\author{Nidhi Majoju}
\orcid{0009-0002-2241-0126}
\email{nidhi.majoju@utdallas.edu}
\affiliation{%
  \institution{University of Texas at Dallas}
  \city{Richardson}
  % \state{Texas}
  \country{USA}
}

\author{Tingxi Li}
\orcid{0009-0005-0975-4428}
\email{tingxi.li@utdallas.edu}
\affiliation{%
  \institution{University of Texas at Dallas}
  \city{Richardson}
  % \state{Texas}
  \country{USA}
}

\author{Aaryaa Moharir}
\email{aaryaa.moharir@utdallas.edu}
\orcid{0009-0000-6717-2040}
\affiliation{%
  \institution{University of Texas at Dallas}
  \city{Richardson}
  % \state{Texas}
  \country{USA}
}

\author{Wei Yang}
\orcid{0000-0002-5338-7347}
\authornote{Corresponding authors.}
\email{yang\_wei@fudan.edu.cn}
\affiliation{%
  \institution{Institute of Systems for Advanced Computing, Fudan University}
  \city{Shanghai}
  % \state{Texas}
  \country{China}
}

\author{Tao Xie}
\orcid{0000-0002-6731-216X}
\authornotemark[2]
\authornote{Tao Xie is also affiliated with the Beijing Tongming Lake Information Technology Application Innovation Center, China, the Institute of Systems for Advanced Computing at Fudan University, China, and the Shanghai Institute of Systems for Open Computing, China.}
\email{taoxie@pku.edu.cn}
\affiliation{%
  \institution{Key Lab of HCST (PKU), MOE; SCS, Peking University}
  \city{Beijing}
  \country{China}
}
% \affiliation{%
%   \institution{Beijing Tongming Lake Information Technology Application Innovation Center}
%   \country{China}
% }

\renewcommand{\shortauthors}{Rathnasuriya, Song, Majoju, Moharir, Li, Yang, and Xie}
% \renewcommand{\shortauthors}{Rathnasuriya and Song, et al.}

%% Abstract
%% Note: \begin{abstract}...\end{abstract} environment must come
%% before \maketitle command
% \begin{abstract}
% \input{tex/abstract}
% \end{abstract}

%% 2012 ACM Computing Classification System (CSS) concepts
%% Generate at 'http://dl.acm.org/ccs/ccs.cfm'.
% <concept_id>10011007.10011006.10011008</concept_id>
% <concept_desc>Software and its engineering~General programming languages</concept_desc>
% <concept_id>10003456.10003457.10003521.10003525</concept_id>
% <concept_desc>Social and professional topics~History of programming languages</concept_desc>
% <concept_significance>300</concept_significance>

% \ccsdesc[500]{Software and its engineering~General programming languages}
% \ccsdesc[300]{Social and professional topics~History of programming languages}
%% End of generated code

%% Keywords
%% comma separated list
%\keywords{Input validation, Program adaptation, Program transformation }  %% \keywords are mandatory in final camera-ready submission

\begin{abstract}

Tile-based programming frameworks are increasingly adopted to write high-performance GPU kernels in domains such as deep learning and scientific computing. While these frameworks enhance productivity and hardware utilization, their multi-stage compilation pipelines introduce distinct code generation bugs that are tightly coupled to input shapes, data types, and backend targets. These bugs often manifest as silent wrong results or performance issues, making them difficult to detect using existing compiler testing tools. Additionally, the unique programming conventions of tile domain-specific languages complicate root cause identification, while fixing such bugs demands specialized knowledge of tile abstractions and compilation pipelines. Despite the growing adoption of tile-based systems, their code generation bugs remain largely unexplored.

This paper presents the first systematic study of tile-program code generation bugs. 
%We curate 401 bug reports from GitHub and identify 301 tile-program codegen bugs for analysis, categorizing the \textit{root causes, symptoms, input patterns, test oracles} that trigger these bugs, and the strategies used to \textit{fix} bugs.
We curate 401 bug reports from GitHub and identify 301 tile-program codegen bugs for analysis, characterizing their \textit{root causes and symptoms}, the \textit{input patterns} that trigger them, the \textit{test oracles} that detect them, and the strategies for \textit{fixing} these bugs.
%We curate 401 bug reports from GitHub and identify 301 tile-program codegen bugs for analysis, categorizing the \textit{root causes, symptoms, input patterns, test oracles} that trigger these bugs and the strategies used to \textit{fix} them. Based on these findings, we develop Tile-CBDetect, a lightweight detection tool that combines automated input generation with differential testing. Tile-CBDetect uncovered nine bugs across 17 operators, across three distinct bug categories. 
Our study provides foundational insights for building debugging, testing, and repair tools tailored to tile-based compiler infrastructures.

%identifying 297 cases 
%linked to code generation bugs. Our study classifies these bugs by root cause and symptom, highlights common triggering conditions, and surveys the repair strategies adopted by developers. Our key findings include: 

%In this paper, we conduct a comprehensive study of GPU-NBs by analyzing 397 real-world bug samples from GitHub. We identify common root causes, symptoms, input patterns, test oracles that trigger these bugs and the strategies used to fix them. 

%We also present GPU-NBDetect, a preliminary tool designed to detect numerical bugs across six distinct bug categories. GPU-NBDetect detected a total of 226 bugs across 186 mathematical functions in four libraries, with 60 of the bugs confirmed by developers. 

\end{abstract}



%% \maketitle
%% Note: \maketitle command must come after title commands, author
%% commands, abstract environment, Computing Classification System
%% environment and commands, and keywords command.
\maketitle

\section{Introduction}

Modern accelerators, including graphics processing units (GPUs)~\cite{nvidiagpus} and artificial intelligence (AI) accelerators~\cite{ahsan2025hardware,cittadini2025hardware,silvano2025survey}, achieve high computational throughput through massive parallelism, wide vector execution, and deep memory hierarchies. Attaining this throughput requires software to coordinate memory access, data locality, work distribution, and thread synchronization. High-level deep learning (DL) frameworks~\cite{paszke2019pytorch,chen2018tvm} and vendor libraries simplify accelerator programming but provide limited control over kernel implementation. In contrast, low-level GPU programming models~\cite{nickolls2007gpu,sanders2010cuda} expose hardware-level control but require programmers to manage scheduling, memory movement, and synchronization manually. These limitations leave a gap between high-level programmability and fine-grained control over kernel execution. 

Tile-based programming frameworks~\cite{openai2021triton,wang2025tilelang,zhang2025hexcute,Nvidia_warp} address this gap by expressing accelerator computations over blocks of data called \textit{tiles}. Each tile represents a logical unit of computation processed cooperatively by a group of threads, whereas the compiler maps the tile-level computation to threads, warps, memories, and accelerator instructions. This separation enables the compiler to optimize data layouts, memory reuse, pipelining, and warp-level execution as the program is lowered through successive intermediate representations (IRs). Tile programming has therefore become a common foundation for generating high-performance kernels on GPUs and AI accelerators.

Although tile programming improves programmability and performance portability, tile-based frameworks translate high-level domain-specific language (DSL) kernels into hardware-specific GPU executables through a multi-stage code-generation (codegen) process~\cite{tillet2019triton,wang2025ml}. During this process, the compiler selects data layouts and schedules, introduces predicates, assigns memory locations, and maps tile operations to device instructions. A defect in the implementation of any of these compiler decisions can propagate through subsequent codegen stages. In this paper, a tile codegen bug is a defect in the codegen process that causes a semantically valid tile program to fail during compilation or produces an executable that violates the program’s intended semantics. Such bugs are difficult to reproduce and diagnose because they often depend on specific tensor shapes, tile configurations, launch configurations, and backend-specific lowering decisions.

Prior work has studied bugs in several compiler and accelerator-programming settings but has not systematically characterized tile codegen bugs. Studies of traditional optimizing compilers analyze general miscompilation bugs~\cite{yang2011finding,livinskii2020random,ma2023survey,even2023grayc}. GPU verification and testing techniques primarily target programmer-written CUDA, OpenCL, and shader programs~\cite{wu2020simulee,barracuda,gklee,gpuverify,boyer2008automated,islam2018bugaroo,sorensen2016exposing,jiang2020cudasmith,binfpe,laguna2019fpchecker,gpu-fpx,laguna2022finding,rathnasuriya2025investigation}. Studies of DL systems and frameworks examine bugs involving operators, APIs, tensor shapes, and backend behavior~\cite{DLGPU1,DLGPU2,DLGPU3,DLGPU4,DLGPU6,DLGPU7}. IR-level testing and fuzzing techniques for tensor compilers target compiler transformations~\cite{limpanukorn2024fuzzing,wang2025duoreduce,wang2023mlirsmith,suo2025desil}. However, prior work has not characterized tile codegen bugs as a distinct class of compiler bugs shaped by tile-specific decisions in layout selection, scheduling, execution mapping, predication, and memory placement. Therefore, prior work does not systematically explain why tile codegen bugs arise, how they manifest, which conditions trigger them, or how developers repair them.  

Tile codegen bugs are difficult to characterize because a bug may be introduced during one compilation stage but become observable only after subsequent lowering stages~\cite{tillet2019triton,openai2021triton}. A bug introduced during a tile-specific transformation can propagate through later stages before producing a compilation failure, runtime failure, or incorrect output~\cite{lattner2008llvm,limpanukorn2024fuzzing,wang2023mlirsmith,suo2025desil}. The resulting symptom may become observable only after subsequent lowering stages and may vary across tensor shapes or compilation configurations. This separation between a bug’s origin and manifestation complicates reproduction and diagnosis, requiring a systematic analysis of root causes, symptoms, triggering conditions, and fixes.

Existing compiler-testing techniques do not directly target the input space and transformation decisions of tile compilation~\cite{yang2011finding,limpanukorn2024fuzzing,livinskii2020random,ma2023survey,wang2023mlirsmith,even2023grayc}. General-purpose program generators such as Csmith~\cite{yang2011finding} and IR-level fuzzers such as MLIR-Smith~\cite{wang2023mlirsmith} primarily generate programs or IR fragments whose execution structure is explicit in the input. They do not systematically vary tile shapes, schedules, layouts, or hardware mappings. GPU-specific fuzzers such as CLsmith~\cite{lidbury2015clsmith}, CUDAsmith~\cite{jiang2020cudasmith}, and DarthShader~\cite{darthshader} exercise accelerator backends through explicit CUDA, OpenCL, or shader programs. Because these tools bypass tile-level abstractions, they provide limited coverage of tile-specific codegen transformations, including multilevel tiling, warp specialization, shared-memory staging, and boundary-mask construction. 

Constructing test oracles for tile compilation is challenging because many tile frameworks lack CPU-executable reference implementations with equivalent kernel semantics. Testing must instead execute the generated GPU kernels and compare their outputs against established baselines, such as vendor libraries or independently developed reference kernels~\cite{cuBLAS,flashmla2025}. Backend-specific lowering may change operation ordering, precision, or instruction selection, while mixed-precision arithmetic and floating-point non-associativity can produce output differences. These effects complicate output-based test oracles because an observed difference may reflect either acceptable numerical variation or a codegen bug that produces incorrect output.

To establish an empirical foundation for detecting, debugging, and repairing tile codegen bugs, we conduct a systematic study across eight open-source tile programming frameworks. We manually examine 401 bug reports and identify 301 as tile codegen bugs. From this corpus, we derive a root-cause taxonomy comprising six categories: \textit{(1) control flow and scheduling, (2) IR construction and transformation, (3) tile mapping and launch, (4) memory, (5) type and operator}, and \textit{(6) device-specific bugs}. We characterize each category by its observable symptoms, bug-triggering input patterns, test oracles, and developer-applied repair strategies. Our findings provide an empirical basis for developing a systematic framework for addressing tile codegen bugs. The framework begins with bug detection (Section~\ref{automated}) through test-input generation and test-oracle selection. The framework then uses the symptoms observed during detection to guide debugging by mapping them to potential root causes (Section~\ref{symptoms}). After identifying the root cause, the framework proceeds to bug repair (Section~\ref{fixing}) by selecting a corresponding repair strategy. 

This study makes the following contributions: 

\begin{itemize}[leftmargin=*]
    \item A curated dataset of 301 real-world tile codegen bugs collected from eight open-source tile programming frameworks.
    \item A root-cause taxonomy comprising six categories, together with a characterization of the relationships among root causes, observable symptoms, and developer-applied repair strategies.
    \item A characterization of bug-triggering input patterns and test-oracle practices, providing empirical guidance for structured input generation and multi-oracle tile compiler testing.
    \item Implications for automated detection, debugging, and repair of tile codegen bugs, organized by compiler stage and tile-specific transformation decision.
\end{itemize}

\section{Background and Motivation}
\label{background}

\begin{figure}[htbp]
\centering
\includegraphics[width=0.99\columnwidth]{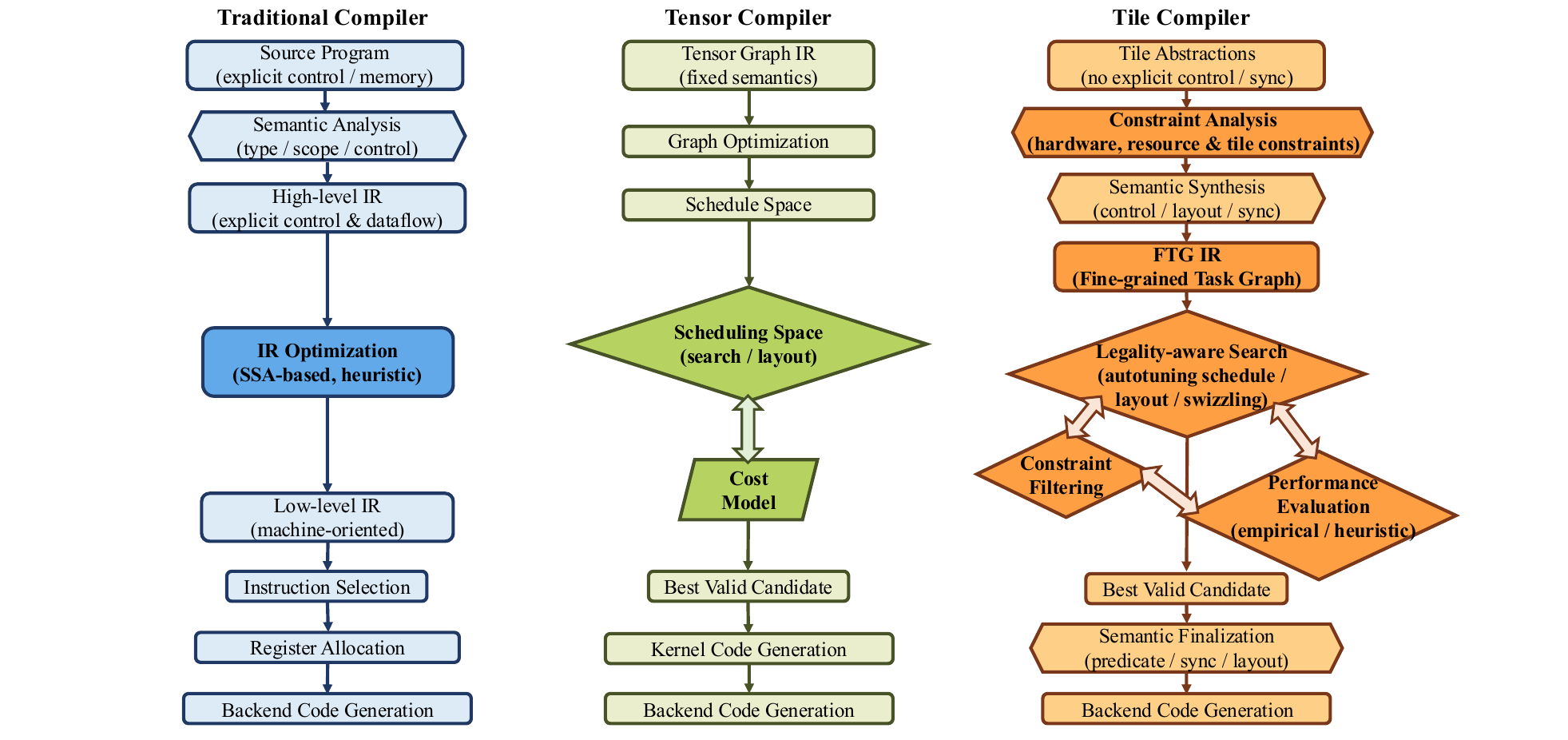}
\caption{Conceptual comparison of compilation pipelines across traditional, tensor, and tile-based compilers.}
\label{fig:compilers}
\end{figure}

GPU kernels often require implementation decisions that are not fixed by the operator definition. Many kernels differ in correctness and efficiency based on how computation is partitioned into tiles, how data is staged across memory levels, and how parallel execution is coordinated. Tile programming~\cite{openai2021triton,wang2025tilelang,zhang2025hexcute,Nvidia_warp} exposes these decisions at the kernel-construction level. Tile compilation then translates tile-level intent into a concrete implementation that satisfies hardware constraints. Fig.~\ref{fig:compilers} summarizes how this pipeline differs from traditional compilation and tensor graph compilation.

\subsection{Tile programs}

A tile program represents a kernel as operations over blocked fragments of tensors. Each program instance operates on one or more tiles mapped to a GPU execution scope, such as a thread block or a warp-aligned group. Compared with tensor graph representations, tile programs make three categories of execution decisions explicit. The first category concerns tile-level data movement, including how values are loaded, staged in registers or shared memory, reused across computation steps, and written back. The second category concerns tile-level parallel mapping, including how program instances cover the iteration space and assign work to threads~\cite{openai2021triton,Nvidia_warp}. The third category concerns layout-sensitive access properties, including indexing patterns, alignment requirements, padding, vectorization opportunities, and explicit layout transformations~\cite{wang2025tilelang}. By making these execution decisions explicit, tile programs expose information relevant to memory access, parallel execution, and execution efficiency while preserving the structured representation needed for compiler analysis and transformation. %By exposing these execution decisions, tile programs provide information relevant to memory access and execution efficiency while preserving a structured representation for compiler analysis and transformation.

%By making these execution decisions explicit, tile programs expose information relevant to GPU execution while preserving the structured representation needed for compiler analysis and transformation.

%\textbf{Tile compilation.} The tile compiler produces a correct and efficient kernel under hardware and resource limits. In contrast to compilation settings where control flow and memory behavior are fully specified by the input program, tile compilation often begins from abstractions that describe tiled computation and dataflow while leaving some kernel details unresolved. The compiler must determine concrete boundary handling, synchronization structure, and layouts, and must also select among many implementation alternatives.

% ISSTA CR \textbf{Tile compilation.} A tile compiler translates a tile program into a hardware-specific GPU kernel. In handwritten CUDA and other imperative GPU programs, the source code explicitly specifies control flow and memory operations. Tile programs instead specify tiled computation and dataflow while leaving boundary handling, synchronization structure, and data layouts to the compiler. The compiler determines these properties under hardware and resource constraints and selects among alternative kernel implementations.

\textbf{Tile compilation.} A tile compiler translates a tile program into a hardware-specific GPU kernel. In handwritten CUDA and other imperative GPU programs, source code explicitly specifies low-level control flow and memory operations. In contrast, tile programs specify tiled computation, dataflow, parallel mapping, and layout-sensitive access properties but leave the hardware-specific realization of these properties to the compiler. Under hardware and resource constraints, the compiler determines concrete boundary predicates, synchronization operations, memory layouts, and instruction mappings.

\subsection{Comparison of Compiler Pipelines}

Fig.~\ref{fig:compilers} contrasts three compilation pipelines and highlights why tile compilation warrants separate study. Traditional compilers lower source programs with explicit control and memory behavior through semantic analysis and IR transformations, then apply backend phases such as instruction selection and register allocation on explicit dependence and memory representations. Tensor graph compilers begin from fixed operator semantics, apply graph rewrites, and search a schedule space (e.g., tiling, fusion, layout) using cost models or measurements before emitting kernels. Tile compilers instead begin with tile abstractions that leave aspects of layout and coordination unresolved; the pipeline performs feasibility checks under resource constraints, synthesizes explicit boundary handling and synchronization, and then searches schedule and layout parameters subject to legality constraints before committing final kernel details. The tile-compilation pipeline therefore introduces bug opportunities beyond conventional lowering and tensor scheduling, including errors in constraint reasoning, boundary and synchronization synthesis, and consistency between legality checks and generated code. Such errors often manifest only under specific combinations of tile sizes, layouts, and resource constraints, making them difficult to expose through conventional compiler or tensor-graph testing. %The tile-compilation pipeline introduces additional bug opportunities beyond conventional lowering and tensor scheduling, including errors in constraint reasoning, boundary and synchronization synthesis, and consistency between legality checks and generated code. These errors may arise only for specific combinations of tile sizes, layouts, and resource usage.

%This structure introduces additional bug opportunities beyond conventional lowering and tensor scheduling, including errors in constraint reasoning, boundary/synchronization synthesis, and mismatches between legality checks and generated code, often triggered only by specific combinations of tile sizes, layouts, and resource usage.

\section{Methodology}
\label{study}

\begin{table*}[ht]
\centering
\caption{Summary of findings (F) and implications (I) in codegen bug study}
\resizebox{\linewidth}{!}{
\begin{tabular}{|p{9cm}|p{10cm}|}
\hline
\textbf{Findings about codegen bug causes} & \textbf{Implications on unique characteristics of tile codegen} \\ 
\hline
\textbf{F1.} Tile program bugs stem from six distinct root causes: control-flow and scheduling, IR construction and transformation, tile mapping and launch, memory, type and operator, and device-specific assumptions (Section~\ref{rootcause} and Table~\ref{tab:root_causes}).  
&
\textbf{I1.} %Unlike traditional workflows where developers explicitly manage memory and control logic, tile DSLs delegate these responsibilities to compiler transformations. Each transformation stage introduces unique failure modes, requiring stage-aware diagnostics. 
Tile program frameworks shift memory and control management to the compiler, introducing distinct failure modes at each transformation stage that demand specialized, stage-aware diagnostics. \\ \hline

\textbf{F2.} Control-flow and scheduling bugs are triggered by incorrect mask generation, instruction reordering, or missed synchronization at tile boundaries (Table~\ref{tab:root_causes}). 
&
\textbf{I2.} Tile DSLs make predication and warp-level synchronization first-class, leading to subtle tile-specific bugs—like silent corruption from miscomputed masks that is not seen in traditional loop transformations. \\ \hline%Tile DSLs elevate predication and warp-level synchronization to first-class constructs. Small mistakes in predicated masking can lead to silent corruption, particularly in partially-filled tiles. These failures have no direct analog in standard loop-level transformations. \\

% \textbf{F3.} IR construction and transformation bugs arise from malformed graphs, incorrect rewrites, or lowered nodes with missing metadata such as type or layout (Table~\ref{tab:root_causes}). 
\textbf{F3.} IR construction and transformation bugs arise from inconsistent construction of tile-aware IR entities, incorrect rewrites, or lowered nodes with missing metadata such as type or layout (Table~\ref{tab:root_causes}). 
&
\textbf{I3.} Tile DSLs extend frameworks like MLIR with custom IRs, but stricter tile-specific semantics make transformations like fusion and splitting prone to brittle index-space misalignment bugs. %Tile DSL compilers often build custom intermediate representations (IRs) on top of existing frameworks like MLIR. However, the semantic contracts for tile-specific rewrites (e.g., fusion, splitting) are stricter, and failures are more brittle due to misalignment in index-space assumptions. \\

\\ \hline

\textbf{F4.} Tile mapping and launch bugs result from incorrect thread-grid calculations, out-of-bound indexing, or misconfigured launch parameters that misalign the kernel’s logical and physical execution dimensions (Table~\ref{tab:root_causes}).
&
\textbf{I4.} Tile DSLs auto-generate launch configs from loop nests, making them vulnerable to subtle errors near tile boundaries or with dynamic shapes—unlike manual control in CUDA. %Unlike handwritten CUDA, where launch configuration is user-controlled, tile DSLs auto-generate launch parameters from high-level loop nests. Bugs in these calculations manifest as subtle correctness errors, especially near boundary tiles or with dynamic shapes. 
\\\hline

\textbf{F5.} Type and operator bugs are among the most frequent causes (48.84\%), often due to unsupported operators, incomplete type inference, or missing type specializations in the backend (Table~\ref{tab:root_causes}).
&
\textbf{I5.} Tile DSLs aggressively specialize for performance but lack standard fallback paths, causing unsupported types or fused ops to silently produce invalid code or incorrect results. %Tile DSLs aggressively specialize code for performance, and lack the fallback mechanisms present in standard libraries. As a result, unsupported types or fused operator patterns can silently generate invalid code or produce incorrect results. 
\\ \hline

%\textbf{F6.} Device-specific bugs occur when generated code includes operations unsupported by certain architectures (e.g., int64 indexing or shared memory overuse), even if syntactically valid.
%&
%\textbf{I6.} Tile DSLs abstract over device capabilities, but do not always reflect hardware constraints during compilation. Without architectural feedback, compilers may emit IR that fails only at runtime on specific GPUs. \\

%\hline 

\textbf{Findings about codegen bug manifestation} & \textbf{Implications on bug detection} \\ \hline

%\textbf{F6.} %Many tile codegen bugs are silent miscomputations or produce undefined behavior under specific inputs, making them difficult to detect via standard exception handling. 
%Many tile codegen bugs result in silent miscomputations or input-specific undefined behavior, escaping detection by standard exception handling. (Section~\ref{oracles})
%&
%\textbf{I6.} Tile codegen bugs often evade runtime checks and lack clear crash signatures, demanding tailored oracles to verify semantic correctness beyond mere execution success.%Unlike traditional compilers with clear crash signatures, tile DSL bugs often bypass runtime checks, requiring tailored oracles that verify semantic correctness beyond execution success. 
%\\ \hline

\textbf{F6.} %Bugs manifest under narrow input regions—specific tensor shapes, data layouts, and numeric values—especially near tile boundaries or for non-divisible extents. 
Bugs emerge under narrow input dimensions like specific shapes, layouts, or values, particularly near tile boundaries or with non-divisible extents (Section ~\ref{inputs}). 
&
\textbf{I6.} %Effective detection requires structured input generation (e.g., around tile multiples, extreme shapes, special values) rather than uniform random fuzzing. Test coverage must account for shape-dependent behavior. 
Effective detection demands structured input generation that targets tile multiples, extreme shapes, and special values since shape-dependent bugs are missed by uniform random manual testing. \\ \hline

\textbf{F7.} Developers use schedule variation, launch sweeps, and cross-device runs to reveal discrepancies, motivating a taxonomy of differential and metamorphic oracles (Section~\ref{oracles}). %Developers often use schedule variation, launch parameter sweeps, or cross-device testing to expose discrepancies, suggesting a natural taxonomy of differential and metamorphic oracles. 

&
\textbf{I7.} %Detection tools must implement semantic-preserving input/output transformations and compare outputs under multiple configurations to catch control-flow, mapping, and device-specific bugs. 
Detection tools should apply semantic-preserving transformations and compare outputs across configurations to expose diverse root causes.\\ \hline

\textbf{F8.} A single output mismatch may stem from diverse root causes such as masking, aliasing, or type issues, therefore no single oracle suffices for all bug types. %A single output mismatch can stem from multiple root causes (e.g., control masking, memory aliasing, type mispromotion), and no single oracle suffices across all bug classes. 
(Section~\ref{oracles})
&
\textbf{I8.} Bug detection requires combining diverse oracles—differential, metamorphic, canary, and algebraic—each targeting specific root causes, underscoring the need for modular, extensible frameworks.
%Bug detection must combine diverse oracle classes—differential, metamorphic, canary injection, and algebraic identities—each tailored to a specific root cause, highlighting the need for modular, extensible testing frameworks. 
\\

\hline

\textbf{Findings about codegen bug symptoms} & \textbf{Implications on debugging} \\ \hline

\textbf{F9.} Crashes are the most frequent symptom, occurring in approximately 58.14\% of the cases, and span multiple stages such as front-end, mid-end, backend, and runtime (Table~\ref{tab:symptoms}).
&
\textbf{I9.} %The phase of the crash (e.g., IR pass vs. runtime) provides valuable diagnostic cues, but developers must correlate stack traces with tile-aware transformation stages for effective triage. 
Crash phases offer diagnostic signals, but effective triage requires mapping stack traces to tile-specific transformation stages.
\\ \hline

\textbf{F10.} Correctness issues such as silent miscomputations or numeric anomalies occur without triggering crashes and are often architecture- or input-dependent (Table~\ref{tab:symptoms}). %(Table~\ref{symptoms})
&
\textbf{I10.} These bugs evade simple detection and demand oracle-based debugging—like differential execution or contract checks—as traditional debuggers miss silent failures. %These bugs evade simple detection and require oracle-based debugging (e.g., differential execution, contract checks), as traditional debuggers offer limited insight into silent failure modes. 
\\ \hline

\textbf{F11.} Performance bottlenecks appear in correct programs and stem from resource overuse or warp-level misbehavior (Table~\ref{tab:symptoms}). 
&
\textbf{I11.} Performance debugging needs profiling and architectural insight—like shared memory and warp scheduling—as issues often stem from masking, memory, or type bugs.
%Debugging performance symptoms requires profiling and architectural awareness (e.g., shared memory usage, warp scheduling), as they often trace back to masking, memory, or type-handling bugs. 
\\ \hline

\textbf{F12.} The mapping between symptoms and root causes is many-to-many (Table~\ref{tab:symptoms}).
&
\textbf{I12.} %Superficial symptoms are insufficient for root cause attribution. Debugging requires cross-phase reasoning about compiler transformations, memory layout, and backend constraints. 
Superficial symptoms rarely reveal root causes; effective debugging requires cross-phase reasoning across compiler transforms, memory layouts, and backend constraints.\\  \hline

\textbf{Findings about codegen bug fixing} & \textbf{Implications on program repair} \\ \hline

\textbf{F13.} Fixes for control and scheduling bugs involve modifying loop guards, predication, and instruction orderings across compiler stages (Table~\ref{tab:fix_totals}). 
&
\textbf{I13.} Fixing control-flow errors requires global reasoning across transformation stages, with attention to data dependencies, mask coverage, and warp divergence.
%Correcting control-flow errors demands global reasoning across transformation stages, requiring awareness of data dependencies, mask coverage, and warp divergence semantics. 
\\ \hline

\textbf{F14.} IR transformation bugs are repaired through low-level rewrites to operator lowering, type emission, and pass composition (Table~\ref{tab:fix_totals}). 
&
\textbf{I14.} %IR-related fixes require fine-grained knowledge of internal representations and backend legality, increasing the repair barrier for non-compiler developers. 
IR-related fixes demand detailed knowledge of internal representations and backend constraints, raising the repair barrier for non-compiler developers. \\ \hline

\textbf{F15.} Fixes for tile mapping and launch bugs adjust kernel tiling parameters and thread-block layouts to restore spatial coverage (Table~\ref{tab:fix_totals}). 
&
\textbf{I15.} %Launch-related repairs often require manual tuning and revalidation across shapes, exposing a gap in abstraction-level consistency and auto-tuning support.
Launch-related fixes often need manual tuning and shape-wise revalidation, revealing gaps in abstraction consistency and auto-tuning support. \\ \hline

\textbf{F16.} Type and operator bugs are fixed by correcting intrinsics, enforcing IEEE-754 semantics, and aligning datatypes with hardware constraints (Table~\ref{tab:fix_totals}). 
&
\textbf{I16.} Precision-sensitive bugs expose missing semantic checks for numerical safety, forcing developers to account for hardware-specific type behavior. %Precision-sensitive bugs reveal a lack of built-in semantic validation for numerical safety, requiring developers to anticipate hardware-specific type behavior. 
\\ \hline

%\textbf{F5.} Device-specific bugs are patched through backend specialization and capability checks targeting ISA or resource mismatches. (See §Fixing Device-Specific)
%&
%\textbf{I5.} Fixes must be hardware-aware and evolve with GPU generations, suggesting the need for architecture-aware static checks and adaptive codegen guards. \\

%\bottomrule

\end{tabular} }
\label{tab:tile-findings}
\end{table*}

%In this section, we describe our data collection process (Section~\ref{data_collection}) and discuss the methodology for filtering and labeling codegen bugs (Section~\ref{labeling}).

This section describes our data collection process (Section~\ref{data_collection}) and the procedure used to filter and label tile codegen bugs (Section~\ref{labeling}).

\subsection{Data Collection}
\label{data_collection}

%To investigate codegen bugs in real-world tile-based programming frameworks, we curated a dataset of verified bug reports from actively maintained open-source repositories. Our goal was to focus on bugs encountered in both research and production usage, ensuring that our analysis reflects the challenges faced by practitioners. The collection process comprises two main stages: (1) identifying relevant repositories and (2) extracting valid bug reports.

To investigate codegen bugs encountered in real-world use of tile-based programming frameworks, we curated a dataset of verified bug reports from actively maintained open-source repositories. We included bugs reported in both research and production settings. The collection process comprised two stages: (1) identifying relevant repositories and (2) extracting valid bug reports.

\subsubsection{Repository Selection}

We leveraged GitHub~\cite{github} as the primary data source due to its central role in hosting tile programming frameworks and its extensive ecosystem of dependent projects. GitHub provides structured metadata for issue tracking, pull request discussions, and code evolution, making it well-suited for empirical analysis of software failures~\cite{Franco,DLGPU6,rathnasuriya2025investigation}.

To improve search precision and reduce irrelevant results, we constructed compound keyword queries based on the terminology commonly used to describe tile programs in documentation and community discussions~\cite{openai2021triton,tillet2019triton}. Specifically, we searched for projects containing terms such as ``tile language'', ``tile kernels'', and ``tile programming'', yielding 155, 45, and 914 repositories, respectively. These targeted queries substantially improved search precision over a naive query using ``tile'' alone, which retrieved over 56,000 repositories, most of which were unrelated, for example, containing references to Triton Inference Server~\cite{nvidia2025triton} or unrelated software sharing the same name.

To finalize the selection, we applied three filtering criteria: (1) \textbf{project activity}: repositories were required to show active development and community interaction including at least 200 GitHub stars, non-trivial issue discussions, and frequent contributions;  (2) \textbf{direct tile usage}: only repositories that implemented tile programs or directly invoked tile-program APIs were retained. Projects using transitive dependencies or wrappers were excluded; (3) \textbf{semantic relevance:} we manually excluded unrelated projects that used the term ``tile'' for other purposes (e.g., graphical tiling, branded inference runtimes) and retained only those built on compiler frameworks that expose tile-level abstractions. This step reduced the number of repositories to 15. 

%After de-duplication and validation, we selected eight repositories that represent both academic and production use of tile-based programming models. These include OpenAI Triton~\cite{Triton}, TileLang~\cite{TileAi}, Halide~\cite{Halide}, TVM (tile-programming components only)~\cite{Apache_tile}, XLA~\cite{Openxla}, DaCe~\cite{Spcl}, NVIDIA Warp~\cite{Nvidia_warp}, and PyTorch (tile-specific modules)~\cite{Pytorch}. 
After removing duplicate repositories and confirming their relevance to tile-based programming, we selected eight repositories used in academic or production settings. These repositories include OpenAI Triton~\cite{Triton}, TileLang~\cite{TileAi}, Halide~\cite{Halide}, TVM (tile-programming components only)~\cite{Apache_tile}, XLA~\cite{Openxla}, DaCe~\cite{Spcl}, NVIDIA Warp~\cite{Nvidia_warp}, and PyTorch (tile-specific modules)~\cite{Pytorch}. For completeness, we also considered two emerging tile-programming frameworks, ThunderKittens~\cite{thunderkittens} and CUDA Tile~\cite{cudatile}, but excluded both frameworks because each had too few publicly confirmed codegen bug reports during our collection window to support reliable analysis. A summary of the analyzed GitHub repositories is available on our project website~\cite{tritonbug2025website}.

%For completeness, we also considered two emerging tile-programming frameworks, ThunderKittens~\cite{thunderkittens} and CUDA Tile~\cite{cudatile}, but excluded them because each project had too few publicly confirmed codegen bug reports during our collection window to support reliable analysis.

\subsubsection{Bug Report Extraction}

%From the selected repositories, we collected all GitHub issues filed between January 2022 and November 2025. This time window reflects the period of active adoption following the public release of key tile-programming frameworks, which became production-ready after mid-2021. Earlier reports were excluded to avoid noise from pre-release instability. 

From the selected repositories, we collected all GitHub issues filed between January 2022 and November 2025. We selected this period because it followed the public release of the examined tile-programming frameworks and covered their broader use in research and production after mid-2021. We excluded earlier reports because they primarily concerned pre-release development and instability.

Our extraction process began with a total of 43,255 GitHub issues across the selected repositories. To identify relevant bug reports, we applied a multi-stage filtering strategy designed to ensure both accuracy and relevance: (1) \textbf{Closed and confirmed issues.}  We filtered for issues marked as closed, under the assumption that resolved issues with discussions or commits are more likely to reflect confirmed bugs. This reduced the dataset to 3,283 issues;  
 (2) \textbf{Keyword and label-based filtering.} To isolate bug-related discussions, we applied both keyword filtering and label matching. Keywords such as ``tile error'', ``tile bug'', ``index'', ``compile'', ``synchronization'', ``warp'', ``OOM'', ``data type'', ``indexing error'', ``compile crash'', ``warp mapping'', ``incorrect IR'', ``fix codegen'', and ``fix'' were used to match issue titles and descriptions. We also retained issues explicitly tagged with GitHub labels such as ``bug'', ``error'', or ``fix''. This step narrowed the candidate set to 892 issues;
(3) \textbf{Fix availability.} To support downstream analysis of bug resolution patterns, we further restricted the dataset to issues that included a documented fix, either in a corresponding pull request, commit reference, or linked resolution discussion. This reduced the set to 579 issues; and (4) \textbf{Deduplication and relevance filtering.} Finally, we manually removed duplicate reports and issues unrelated to tile codegen bugs, such as feature requests and configuration questions. After this filtering, 401 reports remained for manual labeling. %Finally, we manually removed duplicates, redundant reports, and issues that were misclassified or irrelevant (e.g., feature requests or configuration questions). This resulted in a curated set of 401 unique candidate bug reports for manual labeling.
%This multi-phase filtering strategy ensured high precision in our bug corpus, allowing us to focus on well-documented and contextually rich failures that accurately reflect the challenges of developing with the tile programming frameworks. 
The multi-stage filtering process produced a high-precision corpus of well-documented reports for analyzing challenges encountered during the development and use of tile-programming frameworks.

\subsection{Bug Labeling and Categorization}
\label{labeling}

To analyze the characteristics of tile-program codegen bugs, we applied a structured labeling procedure informed by prior taxonomies in GPU and compiler bug studies. We began with established root cause classes including memory bugs~\cite{li2019detecting,tiwari2015understanding,sorensen2016exposing,islam2018bugaroo}, synchronization bugs~\cite{wu2020simulee,li2019detecting,yang2012fixing111,gopalakrishnan2021guarding}, numerical bugs~\cite{yang2012fixing111,rathnasuriya2025investigation} or data-type bugs~\cite{rathnasuriya2025investigation}, and general correctness bugs~\cite{li2019detecting,DLGPU7,DLGPU1,islam2019comprehensive}. These categories have been widely used in prior empirical studies of CUDA, MLIR, and DL-framework bugs, providing a validated foundation for our analysis~\cite{islam2019comprehensive,rathnasuriya2025investigation,tiwari2015understanding,DLGPU1}.

We initiated the labeling process with two authors—both of whom have prior experience conducting empirical studies of GPU-related bugs and have published peer-reviewed work in this domain—independently reviewing each candidate bug report.
This review involved a detailed examination of the issue description, discussion threads, and associated commits or pull requests to assign the most appropriate root cause label. During labeling, we observed that traditional bug taxonomies, especially those from CUDA, were not directly applicable to tile codegen bugs. For example, CUDA synchronization bugs refer to low-level thread divergence or barrier misuse. In tile programs, where warp-level parallelism and implicit scheduling are handled by the compiler, synchronization bugs more commonly arise from mismatches in tile boundaries, warp-id semantics, or shape assumptions across thread groups. To accommodate these shifts, we refined the root cause taxonomy to better reflect the abstraction level and failure modes of tile-based compilers. For instance, we introduced the category of ``warp control bugs'' to capture bugs involving incorrect warp/thread mapping, loop unrolling, or tile ID misalignment. 

To ensure consistency across reviewers, we compared the independently assigned labels and merged equivalent categories where appropriate. Disagreements were resolved through a joint adjudication process. During this process, 151 bug reports were identified as inconsistently labeled. After further review, we excluded 100 reports because the available information was insufficient to determine a root cause or because the reported failure resulted from a high-level user error, such as misuse of the tile API rather than a bug in kernel logic. The review yielded a final set of 301 tile codegen bug reports with identifiable root causes for subsequent analysis.

%After further review, we excluded 100 of them due to insufficient information to determine a root cause or because the issue stemmed from high-level user errors (e.g., misuse of the tile API rather than bugs in kernel logic). This left a final set of 301 tile program codegen bug reports that were clearly attributable to well-defined root causes and suitable for downstream analysis. 

\textbf{Taxonomy Construction.} After finalizing the labeled dataset, we developed a bug taxonomy that associates each root cause category with its observable symptoms and developer-applied fixes. Unlike prior studies in compiler bugs that focus primarily on crash vs. non-crash failures~\cite{yang2011finding,livinskii2020random,ma2023survey,even2023grayc,limpanukorn2024fuzzing,wang2025duoreduce,wang2023mlirsmith,suo2025desil}, our analysis revealed a richer set of bug characteristics in tile programs, including shape mismatches, tile alignment bugs, vectorization regressions, memory scope conflicts, and architecture-dependent miscompilations. To construct the taxonomy, we grouped similar symptoms observed across reports and traced them to the compiler transformation responsible for the bug. Fix strategies were derived from the resolution steps taken in patches, including scheduling changes, index expression rewrites, type propagation fixes, and warp mapping adjustments. 

\begin{table}[tbph!]
\centering
\caption{ The taxonomy of bug causes}
\label{tab:root_causes}
\resizebox{\linewidth}{!}{
\begin{tabular}{p{3cm} | p{3.5cm} |p{9cm} |p{1.0cm} |p{1.0cm}}
\toprule
\textbf{Category} & \textbf{Subcategory} & \textbf{Description} & \textbf{\# Bugs} & \textbf{Ratio} \\
\hline

\multirow{3}{3cm}{Control Flow and Scheduling Bugs} 
& Branch Predication & Incorrect predicates or loop guards cause wrong control-flow decisions. & 10& 3.32\%\\
& \cellcolor[HTML]{EFEFEF} Instruction Scheduling & \cellcolor[HTML]{EFEFEF}  
Instruction reordering violates dependencies, causing unsafe execution. & \cellcolor[HTML]{EFEFEF} 2& \cellcolor[HTML]{EFEFEF} 0.66\%\\
& Warp Control  & Misuse of warp intrinsics causes incorrect thread coordination. & 4& 1.33\%\\

&  \cellcolor[HTML]{EFEFEF}\textbf{Subtotal} & \cellcolor[HTML]{EFEFEF}\textbf{-} &\cellcolor[HTML]{EFEFEF}\textbf{16}&\cellcolor[HTML]{EFEFEF} \textbf{5.32\%}\\

\hline

\multirow{2}{3cm}{IR Construction and Transformation Bugs} 
& IR Construction  & Result from incorrect IR generated during compiler lowering of valid source programs, such as wrong types, shapes, or memory annotations. & 4& 1.33\%\\
& \cellcolor[HTML]{EFEFEF} IR Transformation  &\cellcolor[HTML]{EFEFEF}  Arise from wrong rewrite or optimization passes that violate tile-specific invariants during IR transformation. 
&\cellcolor[HTML]{EFEFEF} 45& \cellcolor[HTML]{EFEFEF} 14.95\%\\
&  \textbf{Subtotal} & \textbf{-} & \textbf{49}& \textbf{16.28\%}\\

\hline

\multirow{2}{3cm}{Tile Mapping and Launch Bugs} 
& Launch Configuration   & Miscomputed launch geometry produces invalid grid, block, or memory sizing. & 5 & 1.66\%\\
& \cellcolor[HTML]{EFEFEF} Thread Block Mapping  & \cellcolor[HTML]{EFEFEF} Wrong mapping logic assigns threads to incorrect data regions. & \cellcolor[HTML]{EFEFEF} 14& \cellcolor[HTML]{EFEFEF} 4.65\%\\

&  \textbf{Subtotal} & \textbf{-} & \textbf{19}& \textbf{6.31\%}\\

\hline

\multirow{3}{3cm}{Memory Bugs} 
& Indexing and Stride  & Incorrect address computation due to invalid strides, broadcasts, or layout transformations. & 35& 11.63\%\\
& \cellcolor[HTML]{EFEFEF} Resource Allocation & \cellcolor[HTML]{EFEFEF} Misestimated buffer sizes, alignments, or memory spaces cause overflows or invalid accesses. & \cellcolor[HTML]{EFEFEF} 16 & \cellcolor[HTML]{EFEFEF} 5.32\%\\
&  Ordering and Caching & Missing synchronization or improper cache usage leads to stale or inconsistent data reads. & 7& 2.33\%\\
&  \cellcolor[HTML]{EFEFEF}\textbf{Subtotal} & \cellcolor[HTML]{EFEFEF}\textbf{-} &\cellcolor[HTML]{EFEFEF} \textbf{58}&\cellcolor[HTML]{EFEFEF} \textbf{19.27\%}\\
\hline

\multirow{3}{3cm}{Type and Operator Bugs} 
& Special Value Handling & Arithmetic fails to correctly propagate or detect NaN, infinities, or denormal values. & 9 & 2.99\%\\
& \cellcolor[HTML]{EFEFEF} Data Type Semantics & \cellcolor[HTML]{EFEFEF} Loss of numeric meaning or precision from implicit casts or inconsistent mixed‑precision handling across devices. & \cellcolor[HTML]{EFEFEF}58 &   \cellcolor[HTML]{EFEFEF} 19.27\%\\
& Operator Implementation & Operator logic is incorrect or incomplete after specialization for type or tile shape.& 80& 26.58\%\\

&  \cellcolor[HTML]{EFEFEF} \textbf{Subtotal} & \cellcolor[HTML]{EFEFEF}\textbf{-} &\cellcolor[HTML]{EFEFEF}\textbf{147}&\cellcolor[HTML]{EFEFEF} \textbf{48.84\%}\\
\hline

Device-Specific Bugs & - & Bugs caused by unsupported or inconsistent backend features across GPU architectures or toolchain layers. & 12& \textbf{3.99\%}\\

\bottomrule
\end{tabular}
}
\end{table}

\section{Bug Causes}
\label{rootcause}

We identified 301 codegen bugs and classified them into six primary categories according to which compiler-synthesized tile semantics are violated: \textit{control flow and scheduling, IR construction and transformation, tile mapping and launch, memory, type and operator,} and \textit{device-specific} bugs. Table~\ref{tab:root_causes} summarizes the distribution of bugs across these categories (I1 in Table~\ref{tab:tile-findings}).

\subsection{Control Flow and Scheduling Bugs}

This category comprises 5.32\% of the observed codegen bugs and captures violations of compiler-synthesized tile control semantics.
Control flow and scheduling bugs arise when the compiler derives control logic—such as boundary guards, predicate masks, and synchronization placement—from high-level tile abstractions, and generates incorrect control behavior during lowering.
In tile compilers, control decisions (e.g., which threads are active, which iterations are valid, and when synchronization is required) are not explicitly programmed but are automatically inferred from the tile iteration space, thread mapping, and shape specialization.
Errors in this synthesis, such as incorrect tile-edge predicates, misgenerated branch conditions, or improper instruction reordering across implicit synchronization points, can violate tile-level execution invariants.
These bugs typically manifest only under specific tile shapes, boundary configurations, or warp layouts, and often remain latent under full-tile execution.
They originate during lowering and scheduling stages, where high-level tile constructs (e.g., loop nests, thread mappings, and boundary guards) are translated into backend control flow and synchronization instructions (I2 in Table~\ref{tab:tile-findings}).
Based on the synthesized semantic object being violated, we identify three subcategories:
\textit{(1) branch predication bugs}, \textit{(2) instruction scheduling bugs}, and \textit{(3) warp-level control bugs}.

\textbf{\textit{Branch Predication Bugs.}}
Branch predication bugs arise when control flow over logical tiles is incorrectly lowered to physical GPU threads.
In tile-based abstractions, constructs such as \texttt{if} statements and boundary guards should predicate only selected logical work items or tile instances without terminating the underlying physical thread.
Incorrect lowering can therefore skip valid tiles or mask computations incorrectly, causing silent loss of work.
In NVIDIA Warp~\cite{warp-issue594}, a conditional \texttt{return} is incorrectly preserved in a CUDA grid-stride loop, causing each physical thread to skip its remaining logical work items.
The fix replaces \texttt{return} with \texttt{continue}, suppressing only the current item.
% https://github.com/NVIDIA/warp/issues/594

\textbf{\textit{Instruction Scheduling Bugs.}}
In tile-based compilers, instruction scheduling bugs arise when backend reordering violates compiler-synthesized tile pipeline invariants, rather than explicit source-level data dependencies.
A representative example from Triton~\cite{triton-issue6750} shows that the AMD backend assumes a fixed per-tile ordering of \texttt{local\_store}, \texttt{global\_load}, compute, and \texttt{local\_load} to implement local prefetching.
A later instruction-reordering pass moves \texttt{global\_load} ahead of \texttt{local\_store}, causing the operations to cross synchronization and barrier boundaries assumed by the local-prefetch scheduler.
% https://github.com/triton-lang/triton/issues/6750

\textbf{\textit{Warp Control Bugs.}}
Warp control bugs in tile-based compilers arise from incorrect synthesis or propagation of warp-level execution metadata, rather than from programmer-written warp logic.
As a result, inconsistencies in warp-related metadata (e.g., mismatches between logical tile layouts and physical warp configurations) can invalidate warp-synchronous assumptions during codegen.
Because warp-related metadata is synthesized and propagated across multiple compiler stages, no single transformation validates global consistency between tile-level layout assumptions and backend warp configurations.
A representative example is reported in Triton~\cite{triton-issue2658}.
TritonGPU requires the product of the dimensions in a blocked layout's \texttt{warpsPerCta} field to equal the module-level \texttt{triton\_gpu.num-warps} attribute.
The \texttt{WSMaterialization} pass violates this invariant by modifying \texttt{num-warps} without updating the tensor layouts contained in the module, thereby producing invalid TritonGPU IR.
The subsequent discussion resolved the inconsistency by representing the number of warp-specialization groups separately while leaving \texttt{num-warps} unchanged.
% https://github.com/triton-lang/triton/issues/2658

\subsection{IR Construction and Transformation Bugs}
This category accounts for 16.28\% of the observed codegen bugs and captures violations of compiler-synthesized tile IR semantics that occur during IR construction and IR rewriting prior to target-specific codegen (I3 in Table~\ref{tab:tile-findings}).
In tile compilers, IRs do not merely encode control flow or data dependencies but explicitly materialize tiling metadata such as per-tile buffers, memory layout descriptors, partial iteration coverage, and masked execution regions. These IR-level objects and invariants are automatically derived from tile shapes, scheduling parameters, and thread mappings, and are central to preserving the semantics of tile-structured execution.
IR-level bugs arise when the compiler incorrectly constructs or transforms these tile-aware IRs, violating semantic invariants that govern how logical tiles map to physical memory and execution resources. 
% Such violations stem from inconsistent synthesis or preservation of implicit tile semantics, rather than from malformed IR or incorrect handling of explicit program constructs.
Such violations stem from inconsistent synthesis or preservation of compiler-generated tile semantics, rather than from defects already present in the input IR or incorrect handling of explicit source-level constructs.
As a result, these failures predominantly arise in tile compilers, whose domain-specific IRs critically depend on consistent construction and transformation of tiling metadata across compiler stages.
Based on the stage at which synthesized IR semantics are violated, we identify two subcategories:
\textit{(1) IR construction bugs} and \textit{(2) IR transformation bugs}.

\begin{figure}[t]
  \centering
  \includegraphics[width=\linewidth]{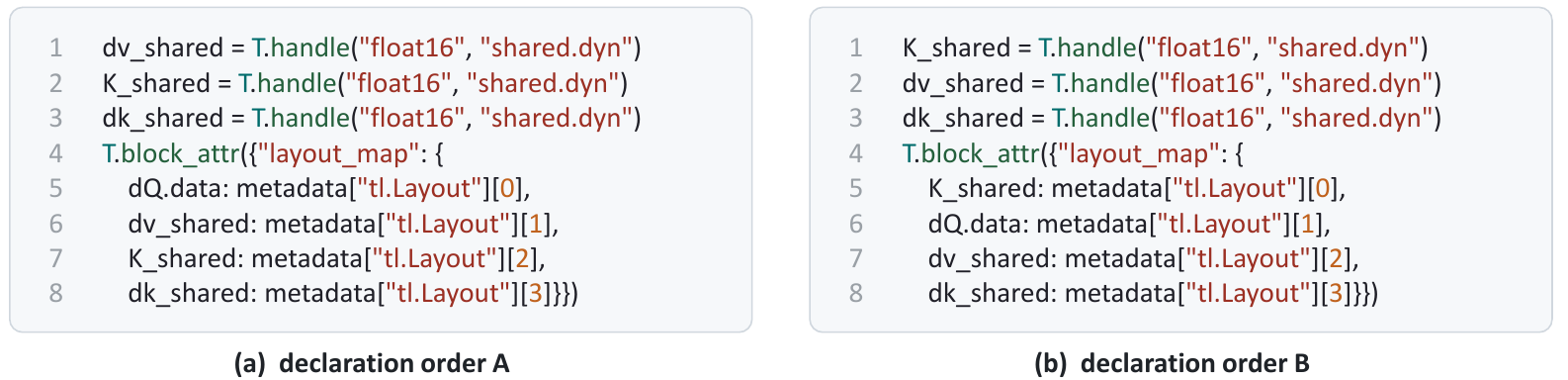}
  \caption{Two serialized IR excerpts reported in TileLang
  issue~\cite{tilelang-issue313}. From left to right, repeated compilation of
  the same FlashAttention kernel emits shared-memory handles and the
  corresponding \texttt{layout\_map} entries in different orders, preventing
  a cache-key match.}
  \label{fig:tilelang-ir-construction}
\end{figure}

\textbf{\textit{IR Construction Bugs.}}
IR construction bugs arise during front-end lowering, where high-level tile programs are translated into tile-aware IRs that explicitly represent per-tile buffers, memory layouts, and layout-to-buffer bindings. At this stage, the compiler must deterministically create and associate IR objects corresponding to shared-memory tiles and their layout metadata.
In tile compilers, these IR objects are first-class and order-sensitive: even when the source program is unchanged, differences in how tile handles or layout mappings are constructed can produce non-canonical serialized IR. IR construction bugs occur when the compiler introduces nondeterminism or inconsistency in the creation or ordering of tile-level IR entities, destabilizing serialization-based comparison and caching.
A representative example is reported in TileLang~\cite{tilelang-issue313}.
Fig.~\ref{fig:tilelang-ir-construction} juxtaposes two serialized excerpts produced by repeated compilations of the same tiled FlashAttention kernel. In the left excerpt, \texttt{dv\_shared} is declared before \texttt{K\_shared}, whereas the right excerpt reverses this order; the corresponding \texttt{layout\_map} entries and metadata indices are reordered consistently with the declarations. Although the resulting kernels are logically equivalent, their serialized IR strings differ, preventing cache-key matching based on \texttt{func.script()}.

% https://github.com/tile-ai/tilelang/issues/313

\textbf{\textit{IR Transformation Bugs.}}
IR transformation bugs arise during mid-end optimization passes that rewrite tile IRs, such as control-flow simplification, predicate rewriting, loop restructuring, padding, or inlining.
These passes often reuse transformation logic originally designed for uniform or affine iteration spaces, implicitly assuming full domain coverage and unconditional execution.
In tile-based IRs, iteration spaces are explicitly structured into tiles with boundary conditions and masked regions.
Transformation bugs occur when IR rewrites invalidate these tile-specific execution invariants, for example, by dropping boundary tiles, simplifying predicates under incorrect assumptions, restructuring control flow in a way that excludes valid tile instances, or invalidating block and producer--consumer relationships.
Such violations may silently reduce the effective iteration domain, generate incorrect code, or produce invalid IR that causes compilation to fail.
An illustrative example is reported in TVM~\cite{tvm-issue16614}.
During multilevel tiling with tensor intrinsics, \texttt{TileWithTensorIntrin} incorrectly applies \texttt{ComputeInline} to blocks not padded by \texttt{PadEinsum}, removing blocks still referenced by the tensorization state and causing schedule construction to fail.
% https://github.com/apache/tvm/issues/16614

\subsection{Tile Mapping and Launch Bugs}

This category accounts for 6.31\% of the observed codegen bugs and captures failures in how tile-level computations are mapped to the underlying GPU execution geometry (I4 in Table~\ref{tab:tile-findings}). Unlike control-flow or IR-level bugs, tile mapping and launch bugs arise at the stage where logical tiles are concretized into physical execution parameters, including grid dimensions, block sizes, and thread--tile associations. Errors in this synthesis can lead to incorrect or invalid kernel launches, even when the source tile program is semantically valid.
We classify tile mapping and launch bugs into two subcategories: \textit{(1) launch configuration bugs} and \textit{(2) thread--block mapping bugs}, depending on whether the failure originates from incorrect launch geometry inference or from erroneous mapping between logical tiles and physical threads.

\begin{figure}[t]
  \centering
  \includegraphics[width=\linewidth]{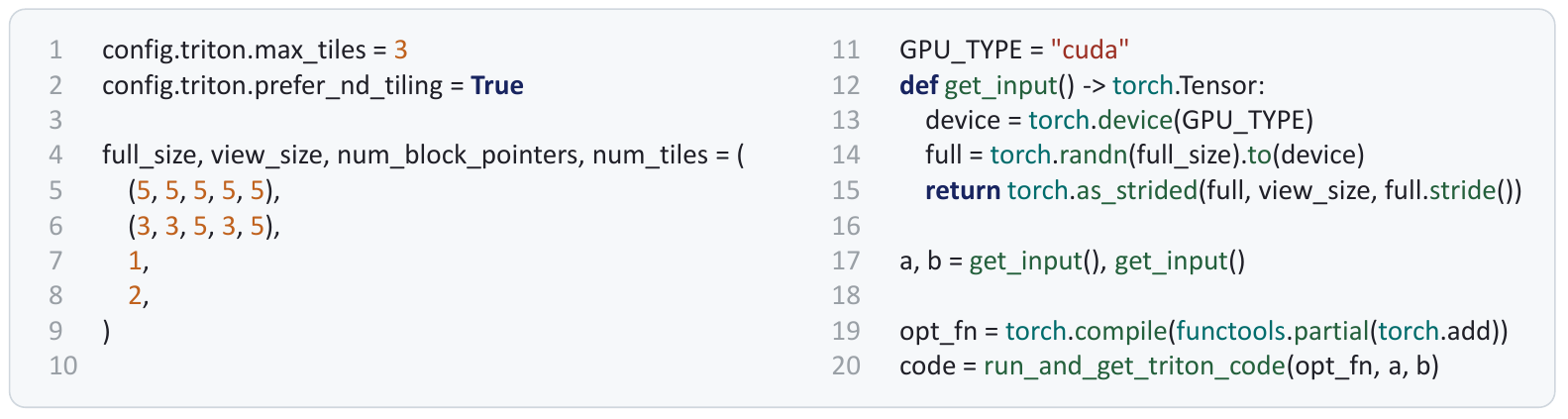}
  \caption{An excerpt from the reproducer in PyTorch
  issue~\cite{pytorch-issue141121}. With three-dimensional tiling enabled,
  the \texttt{z}-prefixed \texttt{LoopBody} iteration variables collide with the
  range-tree symbols for the third tiling axis, triggering an internal
  assertion during code generation.}
  \label{fig:pytorch-three-tile-launch}
\end{figure}

\textbf{\textit{Launch Configuration Bugs.}}
Launch configuration bugs arise when the compiler miscomputes launch parameters such as grid size, block dimensions, or shared memory usage derived from tile-level abstractions. 
Tile compilers infer these parameters from input shapes, tile sizes, and tile grouping constraints. As a result, incorrect assumptions during launch synthesis (e.g., implicit divisibility, warp alignment, or resource availability) can produce invalid or incomplete execution.
A representative example is reported in PyTorch's Triton integration~\cite{pytorch-issue141121}.
Fig.~\ref{fig:pytorch-three-tile-launch} shows the reproducer: with \texttt{config.triton.max\_tiles = 3} and \texttt{config.triton.prefer\_nd\_tiling = True}, Inductor exercises its rarely used three-dimensional tiling path.
Inductor then generates ordinary \texttt{LoopBody} iteration variables with a \texttt{z}-prefixed name that overlaps with the range-tree symbols used for the third tiling axis.
This naming collision causes an internal assertion during code generation, before the kernel is launched.
The fix introduces explicit \texttt{ZBLOCK} support and assigns a distinct prefix to ordinary \texttt{LoopBody} iteration variables.
% https://github.com/pytorch/pytorch/issues/141121

\textbf{\textit{Thread-Block Mapping Bugs.}}
Thread--block mapping bugs arise when the compiler emits incorrect logic for mapping logical tiles to physical grid and block indices. Tile compilers frequently decompose multidimensional iteration spaces (e.g., batch, head, or sequence dimensions) into tiles, and then flatten or distribute these tiles across the limited grid axes supported by GPUs. Errors in this mapping process can lead to invalid grid dimensions, overlapping execution, or incomplete coverage of the logical iteration space.
An illustrative example is reported in PyTorch~\cite{pytorch-issue157018}.
When \texttt{flex\_attention} is compiled with \texttt{dynamic=True}, the generated Triton launch maps the combined batch and head dimensions to the CUDA \texttt{gridY} axis.
For example, $B=22{,}720$ and $H=3$ produce a \texttt{gridY} dimension of $68{,}160$, exceeding CUDA's limit of $65{,}535$ and causing a runtime ``invalid argument'' error.
The corresponding compilation with \texttt{dynamic=False} executes successfully.
This failure therefore results from an inappropriate assignment of logical batch and head tiles to a constrained physical grid axis.
% https://github.com/pytorch/pytorch/issues/157018

\subsection{Memory Bugs}

This category accounts for 19.27\% of all observed codegen bugs and captures failures in translating compiler-synthesized tile memory semantics into correct low-level memory operations (I1 in Table~\ref{tab:tile-findings}). Unlike control-flow or mapping bugs, which govern execution and scheduling, tile memory semantics bugs arise when the compiler incorrectly derives how memory behavior is structured across tiles and tile execution stages.
In tile-based compilers, memory layouts, buffer lifetimes, and synchronization boundaries are derived from tile configurations and scheduling decisions, rather than explicitly specified in the source program. Errors in these derivations can therefore lead to silent data corruption or incorrect results even when the source tile program is semantically correct.
We classify tile memory semantics bugs into three subcategories: \textit{(1) indexing and stride bugs}, \textit{(2) resource allocation bugs}, and \textit{(3) ordering and caching bugs}, based on which aspect of compiler-synthesized memory behavior is violated.

\textbf{\textit{Indexing and Stride Bugs.}}
Indexing and stride bugs arise when the compiler incorrectly synthesizes address computations from logical tile indices or assumes incompatible memory strides during tile lowering. Tile programs typically express memory access in terms of tile-local indices, while the compiler is responsible for lowering these accesses into concrete global or shared memory operations, including stride calculation, boundary masking, and tail handling. Errors in this synthesis can cause incorrect memory accesses even though the source program does not explicitly perform invalid indexing.
A representative example is reported in
Triton~\cite{triton-issue832}, where a valid flattened index exceeding \texttt{INT\_MAX} is computed using signed 32-bit arithmetic. The generated index computation overflows, producing an invalid address and an illegal memory access.
% https://github.com/triton-lang/triton/issues/832

\begin{figure}[t]
  \centering
  \includegraphics[width=\linewidth]{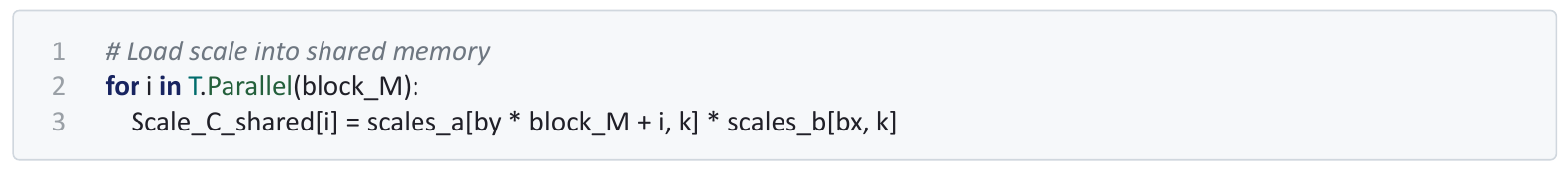}
  \caption{The scale-loading operation from TileLang issue~\cite{tilelang-issue359}. The \texttt{T.Parallel} assignment is lowered to a producer-warpgroup SIMT copy that requires intermediate registers, but automatic register deallocation conflicts with the need for intermediate registers during this copy.}
  \label{fig:tilelang-resource-allocation}
\end{figure}

\textbf{\textit{Resource Allocation Bugs.}}
Resource allocation bugs arise when the compiler incorrectly infers the lifetime or reuse of registers or shared-memory buffers across tile phases. Tile compilers aggressively optimize memory usage by overlapping tile computation stages, reusing registers, and minimizing shared-memory footprints. These optimizations rely on precise lifetime analysis across tile-parallel regions. When such analysis is incorrect, registers or buffers may be freed or reused prematurely, leading to incorrect computation.
An illustrative example is reported in TileLang~\cite{tilelang-issue359}, where a tiled DeepGEMM kernel deadlocks on an NVIDIA Hopper GPU when configured with eight warps and warp specialization.
Fig.~\ref{fig:tilelang-resource-allocation} shows the relevant \texttt{T.Parallel} loop, in which each iteration reads scaling values from global memory, multiplies them, and writes the result to the shared-memory tile \texttt{Scale\_C\_shared}. TileLang lowers this direct assignment to a SIMT copy executed by the producer warpgroup. Although the resulting global-to-register-to-shared transfer requires intermediate registers, the compiler automatically applies \texttt{warpgroup\_reg\_dealloc} to reduce the producer warpgroup's register allocation.
The maintainer attributed the deadlock to this register-allocation conflict and suggested using \texttt{T.NoSetMaxNReg()} to prevent the automatic register deallocation.
% https://github.com/tile-ai/tilelang/issues/359

\textbf{\textit{Ordering and Caching Bugs.}}
Ordering and caching bugs arise when the compiler mismanages tile-derived ordering or reuse assumptions during codegen. Tile compilers often emit synchronization primitives and caching decisions implicitly as part of pipelined tile schedules, rather than relying on fixed procedural barriers. Errors in this process can lead to stale data reads, missing memory fences, or reuse of incorrect cached results.
A representative example is reported in NVIDIA Warp~\cite{warp-issue639}. The Cholesky implementation supplied by cuMathDx requires the CUDA block dimension at compile time, but Warp's kernel hash omitted \texttt{block\_dim}. Consequently, launching the same tiled kernel with a different block dimension could reuse an LTO artifact compiled for an earlier launch configuration.
% https://github.com/NVIDIA/warp/issues/639

\subsection{Type and Operator Bugs}
This category accounts for 48.84\% of all observed codegen bugs and captures failures in preserving numeric and operator semantics during tile-level lowering and specialization (I5 in Table~\ref{tab:tile-findings}). 
These bugs arise when high-level arithmetic and storage operations are lowered through vectorization, packing, software emulation, reductions, or target-specific intrinsics.
Incorrect lowering can change a value's representation or violate an operator's semantics even when the source computation is type-correct.
We classify type and operator bugs into three subcategories: 
\textit{(1) special-value handling bugs}, \textit{(2) data-type semantics bugs}, and \textit{(3) operator implementation bugs}.

\textbf{\textit{Special-Value Handling Bugs.}}
Special-value handling bugs occur when the compiler mishandles floating-point edge cases such as NaN propagation, infinities, denormal values, or signed zeros during tile-level execution.
These values may require bit-level distinctions and operator-specific semantics that cannot be recovered from ordinary numerical comparisons.
Incorrect lowering or software emulation can therefore violate operator semantics even when the high-level computation is well-defined.
A representative example is reported in Triton~\cite{triton-issue6376}, where the software implementation of floating-point \texttt{atomic\_max} uses a numerical comparison to determine the sign of its operand.
Because \texttt{-0.0} compares equal to \texttt{+0.0} despite having its sign bit set, it is processed through the wrong atomic path and produces an incorrect result.
The fix determines the sign directly from the floating-point bit representation.
% https://github.com/triton-lang/triton/issues/6376

\textbf{\textit{Data-Type Semantics Bugs.}}
Data-type semantics bugs arise when the compiler incorrectly propagates, converts, or represents data types during tile lowering.
Because tile values may be vectorized, packed, and exchanged with framework tensors, such errors can alter precision or bit-level representations even when the source computation is type-correct.
In Triton~\cite{triton-issue756}, boolean \texttt{true} is stored as \texttt{255}, while PyTorch represents it as \texttt{1}, producing inconsistent results when the output is reinterpreted as \texttt{uint8}.
The fix normalizes boolean values before storing them, preserving the framework-visible representation.
% https://github.com/triton-lang/triton/issues/756

\begin{figure}[t]
  \centering
  \includegraphics[width=\linewidth]{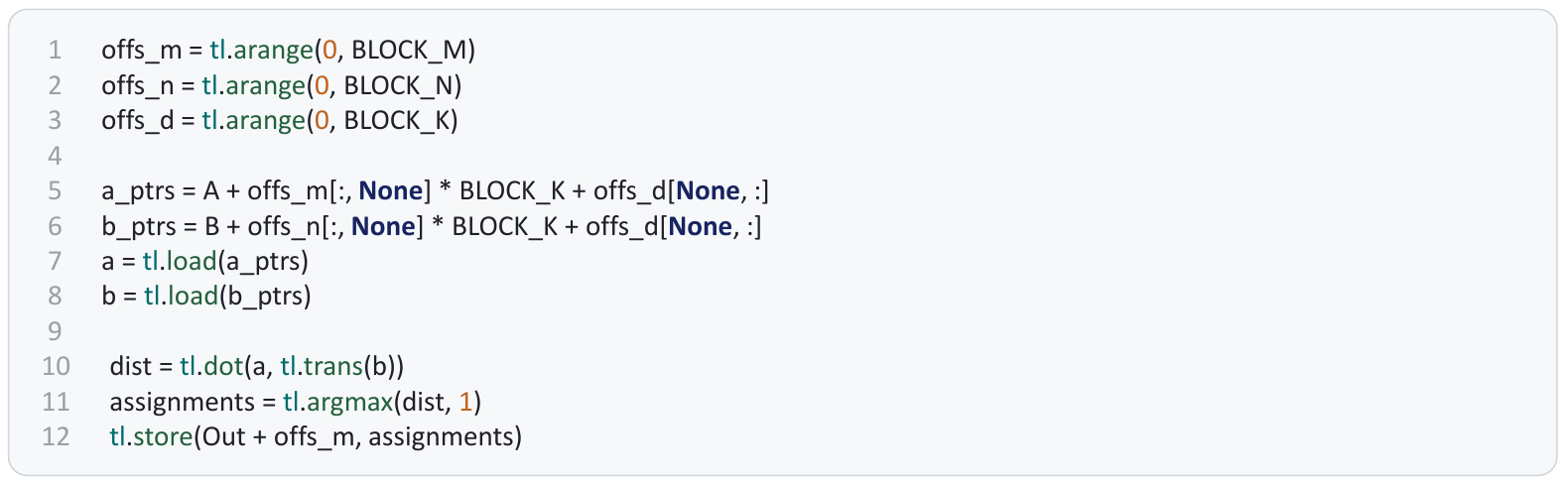}
  \caption{A kernel-body excerpt from Triton
  issue~\cite{triton-issue1846}. A matrix-multiplication tile followed by an
  \texttt{argmax} reduction triggers a compiler segmentation fault during
  layout-conversion removal.}
  \label{fig:triton-argmax-bug}
\end{figure}

\textbf{\textit{Operator Implementation Bugs.}}
Operator implementation bugs arise when the compiler synthesizes incorrect logic for operators under tiling, masking, or fusion. In tile-based compilers, high-level operators are frequently lowered into tile-local algorithms whose control flow, reduction structure, and boundary handling are derived from tile shapes and execution strategies. Errors in this synthesis can violate operator semantics even when the source operator itself is well-defined.
A representative example is reported in Triton~\cite{triton-issue1846}, where a tile-level \texttt{argmax} reduction following matrix multiplication causes the compiler to crash.
Fig.~\ref{fig:triton-argmax-bug} shows the relevant kernel body: the pointer grids select two input tiles, \texttt{tl.dot} produces the tile-local distance matrix, and \texttt{tl.argmax} reduces each row to an index that is stored in \texttt{Out}.
Although both operators are individually valid, their composition triggers a segmentation fault in the operation folder invoked during layout-conversion removal.
% https://github.com/triton-lang/triton/issues/1846

\subsection{Device-Specific Bugs}

This category accounts for 3.99\% of the observed codegen bugs and captures failures caused by incompatibilities between compiler-generated tile kernels and target GPU hardware constraints (I1 in Table~\ref{tab:tile-findings}). 
In these cases, the tile program is semantically valid, but target-specific lowering either imposes an unsupported hardware assumption or fails to represent a configuration supported by the target device.
Such bugs may surface during target-specific compilation or at runtime.

One example is reported in TileLang~\cite{tilelang-issue101}, where an MMA kernel fails during device-specific layout decomposition on an NVIDIA H100. 
The logical tile corresponds to a per-warpgroup MMA shape of $64 \times 96 \times 32$ along the $M$, $N$, and $K$ dimensions, respectively. The reporter expects the underlying WGMMA instructions to support this shape.
However, the selected CuTe lowering path attempts to divide dimensions 96 and 64 and rejects them as non-divisible. Thus, the compiler imposes a layout-divisibility assumption that is incompatible with this otherwise valid hardware-level MMA configuration.
\section{Mapping Symptoms to Root Causes}
\label{symptoms}
 
This section examines how the observed symptoms of tile codegen bugs relate to the six root-cause categories. We classify the symptoms into \textit{crashes, correctness issues}, and \textit{performance bottlenecks}, with the corresponding subcategories shown in Table~\ref{tab:symptoms}. This classification provides a structured basis for tracing observed symptoms to their associated root-cause categories and supports systematic diagnosis of tile codegen bugs.

\begin{table}[htbp]
% \vspace{1mm}
\footnotesize
\caption{Distribution of bug symptoms across causes. Each cause (RC1--RC6) represents a major bug category: RC1 for control-flow and scheduling bugs, RC2 for IR construction and transformation bugs, RC3 for tile mapping and launch bugs, RC4 for memory bugs, RC5 for type and operator bugs, and RC6 for device-specific bugs.}
\label{tab:symptoms}
\resizebox{\linewidth}{!}{
\begin{tabular}{c|c|c|c||cccccc}

% \toprule
% \multicolumn{2}{c|}{\textbf{Symptoms}}        & \multicolumn{2}{c||}{\textbf{Occurrence}} & \begin{tabular}[c]{@{}c@{}}\textbf{RC1}\end{tabular} & \begin{tabular}[c]{@{}c@{}}\textbf{RC2}\end{tabular} & \begin{tabular}[c]{@{}c@{}}\textbf{RC3}\end{tabular} & \begin{tabular}[c]{@{}c@{}}\textbf{RC4}\end{tabular} & \begin{tabular}[c]{@{}c@{}}\textbf{RC5}\end{tabular} & \begin{tabular}[c]{@{}c@{}}\textbf{RC6}\end{tabular}\\ \hline
\toprule
\multicolumn{2}{c|}{\multirow{2}{*}{\textbf{Symptoms}}}
& \multicolumn{2}{c||}{\textbf{Occurrence}}
& \multicolumn{6}{c}{
    \textbf{Root Cause} \textit{(\% within symptom category)}
}\\
\multicolumn{2}{c|}{}
& \multicolumn{2}{c||}{\textit{(\% of all 301 bugs)}}
& \textbf{RC1}
& \textbf{RC2}
& \textbf{RC3}
& \textbf{RC4}
& \textbf{RC5}
& \textbf{RC6}\\
\hline

\multirow{4}{*}{\textbf{Crash}}              & Front-end errors    & 89 (29.57\%) & 
\multirow{4}{*}{\begin{tabular}[c]{@{}c@{}}175\\ (58.14\%)\end{tabular}} & 
\multirow{4}{*}{\begin{tabular}[c]{@{}c@{}} 12\\ (6.86\%)\end{tabular}
} & 
\multirow{4}{*}{\begin{tabular}[c]{@{}c@{}}37\\ (21.14\%)\end{tabular}} & \multirow{4}{*}{\begin{tabular}[c]{@{}c@{}}8\\ (4.57\%)\end{tabular}}  & 
 \multirow{4}{*}{\begin{tabular}[c]{@{}c@{}}32\\ (18.29\%)\end{tabular}}  &
 \multirow{4}{*}{\begin{tabular}[c]{@{}c@{}}\cellcolor[HTML]{EFEFEF}\textbf{76}\\ \cellcolor[HTML]{EFEFEF}\textbf{(43.43\%)}\end{tabular}}  &
\multirow{4}{*}{\begin{tabular}[c]{@{}c@{}}10\\ (5.71\%)\end{tabular}} \\

  & Mid-end errors & 20 (6.64\%) &                          \\
  & Back-end errors & 34 (11.30\%) &                          \\                  
  & In-kernel runtime errors          & 32 (10.63\%)  &  &  & &  & & &\\ \hline

\multirow{6}{*}{\textbf{Correctness issues}} & Silent wrong result                & 23 (7.64\%) & 

\multirow{6}{*}{\begin{tabular}[c]{@{}c@{}} 109\\ (36.21\%)\end{tabular}} &  
\multirow{6}{*}{\begin{tabular}[c]{@{}c@{}}4\\ (3.67\%)\end{tabular}} & 
\multirow{6}{*}{\begin{tabular}[c]{@{}c@{}}10\\ (9.17\%)\end{tabular}} & 
\multirow{6}{*}{\begin{tabular}[c]{@{}c@{}}9\\ (8.26\%)\end{tabular}} & 
\multirow{6}{*}{\begin{tabular}[c]{@{}c@{}}19\\ (17.43\%)\end{tabular}} & 
\multirow{6}{*}{\begin{tabular}[c]{@{}c@{}}\cellcolor[HTML]{EFEFEF}\textbf{67} \\ \cellcolor[HTML]{EFEFEF}\textbf{(61.47\%)}\end{tabular}} & 
\multirow{6}{*}{\begin{tabular}[c]{@{}c@{}}- \\ \end{tabular}} \\

& Run-to-run variance & 2 (0.66\%) &                          \\
& Cross-implementation mismatch & 6 (1.99\%) &                          \\
& Contract violation & 19 (6.31\%) &                          \\
& Wrong buffer/region & 4 (1.33\%) &                          \\
    & Numeric pathology                   & 55 (18.27\%)  &  &  & &  &  & &    \\  \hline

\multirow{3}{*}{\textbf{Performance bottlenecks}}     & Execution hang          & 8 (2.66\%)  & 
\multirow{3}{*}{\begin{tabular}[c]{@{}c@{}} 17 \\(5.65\%)\end{tabular}} & 
\multirow{3}{*}{\begin{tabular}[c]{@{}c@{}}-\\\end{tabular}} & 
\multirow{3}{*}{\begin{tabular}[c]{@{}c@{}}
2\\(11.76\%)\\\end{tabular}} &
\multirow{3}{*}{\begin{tabular}[c]{@{}c@{}}2\\ (11.76\%)\end{tabular}} & 
\multirow{3}{*}{\begin{tabular}[c]{@{}c@{}}\cellcolor[HTML]{EFEFEF}\textbf{7}\\ \cellcolor[HTML]{EFEFEF}\textbf{(41.18\%)}\end{tabular}} & 
\multirow{3}{*}{\begin{tabular}[c]{@{}c@{}}\cellcolor[HTML]{EFEFEF}\textbf{4}\\ \cellcolor[HTML]{EFEFEF}\textbf{(23.53\%)}\end{tabular}} & 
\multirow{3}{*}{\begin{tabular}[c]{@{}c@{}}
2\\(11.76\%)\\ \end{tabular}} \\

& Memory inefficiency & 4 (1.33\%) &                          \\

& Type inefficiency        & 5 (1.66\%)  &  &&&& & &  \\ \bottomrule

\end{tabular}}%}
\end{table}

Crashes are the most frequent symptom, accounting for 175 bugs (58.14\%). Crashes occur when compilation or kernel execution terminates abnormally because of an exception, segmentation fault, device-side error, or kernel-launch failure. They may arise during front-end processing, mid-end transformation, back-end lowering, or in-kernel execution. Type and operator bugs (RC5) constitute the dominant root-cause category, followed by IR construction and transformation bugs (RC2) and memory bugs (RC4). The associated error messages often provide meaningful signals about the responsible transformation phase. For example, front-end messages such as \textit{``Cannot infer type of \texttt{tl.where}''} and \textit{``Mismatched operand types for fusion''} may indicate type and operator bugs or IR construction and transformation bugs. Errors produced during tiling, vectorization, or lowering may indicate control flow and scheduling bugs or tile mapping and launch bugs. For example, \textit{``Expected mask of shape (X, Y), got (X, 1)''} indicates an inconsistency introduced during tile transformation, whereas \textit{``Shared-memory allocation exceeds device limits''} indicates a memory or device-specific constraint violation. Runtime errors, such as invalid memory accesses and CUDA launch errors, may indicate memory bugs, tile mapping and launch bugs, or device-specific bugs. Therefore, crash-related error messages can help localize a bug to a particular phase of the tile compilation pipeline (I9 in Table~\ref{tab:tile-findings}).

Unlike crashes, correctness issues arise when the generated kernel executes to completion but produces an incorrect result. We observe this symptom category in 109 bugs (36.21\%). These symptoms include silent wrong results, numeric pathologies, cross-implementation mismatches, and contract violations. Type and operator bugs (RC5) constitute the dominant root-cause category, followed by memory bugs (RC4). Type and operator bugs can cause precision loss or accumulation drift that appears only under specific data types or fusion patterns. For example, comparing \texttt{float16} and \texttt{float32} executions of the same computation can expose data-type-dependent discrepancies. Memory bugs can produce incorrect values or output shapes when indexing or buffer aliasing behaves differently across tensor layouts. Control flow and scheduling bugs (RC1) may leave partial tiles incorrectly computed or masked, whereas tile mapping and launch bugs (RC3) may duplicate or omit output elements under particular grid or block configurations. Oracle-based checks expose these symptoms by comparing a tile-program kernel with a trusted PyTorch baseline and by varying data types, tile shapes, and launch configurations. Therefore, successful kernel execution does not establish correctness; developers must compare the generated output with a trusted reference implementation or verify that the output satisfies a known metamorphic relation (I10 in Table~\ref{tab:tile-findings}). 

In contrast to crashes and correctness issues, performance bottlenecks arise when a kernel either produces correct results inefficiently or fails to terminate within an expected time bound. We observe this symptom category in 17 bugs (5.65\%), including execution hangs, memory inefficiencies, and type inefficiencies. Profiling tools such as Nsight Compute and PyTorch timers expose them through prolonged execution, excessive shared-memory use, or low hardware utilization. Memory bugs (RC4) constitute the dominant root-cause category, followed by type and operator bugs (RC5). For example, uncoalesced memory accesses and excessive shared-memory allocation can increase data-movement costs. Unnecessary upcasting or incorrect fusion of precision modes can reduce compute throughput. Incorrect use of warp-level intrinsics under predication can serialize execution, and vectorization across partial tiles can introduce shape-dependent inefficiencies. Therefore, runtime measurements and hardware-counter profiles are necessary to distinguish inefficiencies in memory behavior, operator lowering, and tile-to-hardware mapping (I11 in Table~\ref{tab:tile-findings}).

\section{Bug Detection Strategies}
\label{automated}

Based on the findings in Table~\ref{tab:tile-findings}, our study identifies two key components for detecting tile codegen bugs:
\textbf{test input generation} and \textbf{test oracles}. Test input generation aims to trigger bugs through \textit{predefined general inputs} and \textit{feedback-driven customized inputs}. Test oracles then identify bugs using comparison-based methods (I6 in Table~\ref{tab:tile-findings}).

\subsection{Test Input Generation}
\label{inputs}

Our findings suggest that fuzzing can systematize the ad hoc
input-generation practices that developers use to expose codegen bugs. Developers often begin with predefined inputs and randomly modify them to generate additional test cases. The manual and ad hoc nature of these modifications limits coverage and reproducibility. We organize the automation of this practice into two complementary strategies: \textbf{predefined general inputs} and \textbf{feedback-driven customized inputs} (F6 and I6 in Table~\ref{tab:tile-findings}).

\subsubsection{Predefined General Inputs.}

Predefined general inputs systematically vary common tile-program
parameters at points where codegen behavior may change. These inputs cover tensor shapes, memory layouts, tile and launch parameters, and data types and values. \textit{Tensor shapes} should include small tensors, singleton dimensions, tile boundaries, and non-divisible sizes. For a tile size of 32, dimensions 31, 32, and 33 cover the cases immediately below, at, and above the boundary. Prime or non-divisible dimensions can reveal errors in index calculation and grid coverage. \textit{Memory-layout inputs} should preserve the tensor values without preserving the original storage arrangement. A non-contiguous view tests stride handling. A transposed view changes the mapping between logical indices and memory addresses. An offset view can reveal errors in address calculation or alignment. \textit{Tile and launch parameters} should vary the mapping from logical tiles to hardware resources. Changes to the tile size or warp count can reveal synchronization and shared-memory errors. Changes to the vector width or unroll factor can reveal errors in generated control flow. \textit{Data-type and value variations} should check whether the compiler preserves the intended numerical semantics. Mixed-precision inputs test promotion and casting rules. NaNs and infinities test whether special values propagate as required. Signed zeros and values near the limits of a data type can reveal floating-point, overflow, and underflow errors.

\subsubsection{Feedback-Driven Customized Inputs.}

Feedback-driven customized inputs extend predefined general inputs through iterative mutation until the test oracle detects a bug-triggering input. A mutation engine modifies one input characteristic at a time. The engine may change a tensor dimension or data type. It may instead adjust the value range or memory layout. After each mutation, the test oracle checks whether the generated input exposes a codegen bug. Greybox fuzzing~\cite{greybox} can use the oracle result to guide subsequent mutations.

\subsection{Test Oracles}
\label{oracles}

Test oracles identify codegen bugs through comparison-based methods. Our findings show that developers first compare the output of a tile program with that of a corresponding PyTorch program using the same predefined input. If the outputs differ, developers apply customized inputs to determine whether the discrepancy persists. They may also compare alternative kernel configurations or reference implementations to further investigate the bug (F7 and I7 in Table~\ref{tab:tile-findings}). We systematize these ad hoc practices into a test-oracle framework that organizes comparison strategies according to the root-cause categories they target, as summarized in Table~\ref{tab:test-oracles}.

Control-flow and scheduling bugs can be detected through \textit{schedule-invariance}, \textit{padding-equivalence}, and
\textit{canary-tail} checks. Schedule-invariance tests vary the warp count or unroll factor and compare the outputs because these scheduling choices should preserve the kernel semantics. Padding-equivalence tests pad an input with neutral values and compare the cropped result with the output for the original input. Canary-tail checks place sentinel values beyond the valid tile boundary. A modified sentinel indicates an out-of-bounds write.

IR construction and transformation bugs can be detected through \textit{differential testing} and \textit{IR assertions}. Differential testing compiles the same kernel with different pass sequences or optimization levels and compares the outputs. Each compilation path should preserve the kernel semantics. IR assertions check whether the generated representation satisfies compiler invariants, including dominance, type consistency, and predicate coverage.

Tile mapping and launch bugs can be detected through \textit{launch-equivalence}, \textit{single-writer}, and \textit{padding-crop} checks. Launch-equivalence tests vary the grid and block factorization while preserving the logical computation. The resulting executions should write the same output elements and produce the same values. Single-writer checks verify that each output index receives exactly one write. Padding-crop tests compare the original output with the cropped output of a padded execution. A mismatch indicates incorrect index mapping or partial-tile handling.

Memory-related bugs can be detected through \textit{layout comparisons}, \textit{canary-region checks}, and \textit{repeated execution}. Layout comparisons run the same computation on contiguous and strided tensors that contain the same values. They can also compare aligned tensors with equivalent offset views. Canary-region checks place sentinel values around an output buffer to detect out-of-bounds writes. Repeated execution with the same input can reveal nondeterministic outputs caused by data races or missing synchronization.

Type and operator bugs can be detected through \textit{precision comparisons} and \textit{special-value checks}. Precision comparisons execute the same operation with different operand or accumulation types and compare the results under an operator-specific tolerance. Special-value checks determine whether NaNs, infinities, and signed zeros propagate according to the operator specification. An oracle can also compare an operator with a semantically equivalent decomposition when both formulations have a defined numerical tolerance.

Device-specific bugs can be detected through \textit{cross-architecture} and \textit{feature-toggle comparisons}. Cross-architecture testing executes the same kernel and input on different GPU architectures or driver versions. An output difference beyond the permitted numerical tolerance may indicate a device-dependent bug in lowering or code emission. Feature-toggle testing (e.g., fast-math) compares executions with an optional hardware feature enabled and disabled. The comparison checks whether the enabled path satisfies the numerical requirements of the kernel.

 \begin{table*}[t]
\centering
\caption{Manifestation strategies and assertions for detecting codegen bugs. Each cause (RC1–RC6) represents a major bug category: RC1 for control-flow and scheduling bugs, RC2 for IR construction and transformation bugs, RC3 for tile mapping and launch bugs, RC4 for memory bugs, RC5 for type and operator bugs, and RC6 for device-specific bugs. }
\resizebox{\textwidth}{!}{
\begin{tabular}{p{2cm}|p{3.9cm}|p{5.8cm}|p{6.1cm}|p{2.2cm}}
\hline
\textbf{Technique} & \textbf{Oracle Instantiation} & \textbf{Manifestation Strategy} & \textbf{Assertion} & \textbf{Root Cause(s)} \\ \hline

% \multirow{3}{2cm}{Differential Testing} 
%  & Schedule variation & Execute same operator under different warp, unroll, or scheduling parameters & Outputs must match; divergence indicates missing barriers or faulty masks & RC1 \\  \cline{2-5}
%  & Pass-pipeline variation & Use alternative compiler pipelines or toggle optimization passes & Outputs remain equivalent; IR must remain verifiable and structurally consistent & RC2 \\\cline{2-5}
%  & Cross-device or feature toggle & Run across architectures or math modes (e.g., TF32, fast-math) & Outputs match within tolerance; deviation reveals backend specialization faults & RC6 \\ \hline

\multirow{3}{2cm}{Differential Testing}
& Schedule variation & Execute the same kernel under different warp, unroll, or scheduling parameters & Outputs must match; mismatches reveal missing barriers or incorrect predication & RC1 \\ \cline{2-5}
& Pass‑pipeline variation & Compile with alternative optimization or transformation pass combinations & Outputs and IR structure remain consistent; deviations indicate invalid transformations & RC2 \\ \cline{2-5}
& Cross‑device / feature toggle & Run identical kernels across architectures or math modes (e.g., TF32, fast‑math) & Results match within tolerance; divergences expose backend or device‑specific bugs & RC6 \\ \hline

% \multirow{3}{2cm}{Metamorphic Testing} 
%  & Padding–cropping & Pad input tensors to tile multiples and crop outputs back & Cropped output equals baseline; reveals partial-tile or boundary errors & RC1, RC3 \\ \cline{2-5}
%  & Launch configuration variation & Vary grid and block sizes with equal total work & Output equivalence; mismatch indicates launch or mapping miscalculation & RC3 \\ \cline{2-5}
%  & Accumulator width sweep & Run same operator with different compute and accumulator precisions & Results within numeric tolerance; large deviation shows type or cast errors & RC5 \\ \hline

 \multirow{3}{2cm}{Metamorphic Testing}
& Padding–cropping & Pad inputs to tile-aligned sizes and crop outputs to original shape & Cropped outputs must match baseline; detects partial-tile or masking errors & RC1, RC3 \\ \cline{2-5}
& Launch configuration variation & Change grid/block tilings with equivalent iteration space & Outputs must match; mismatches reveal launch or mapping bugs & RC3 \\ \cline{2-5}
& Accumulator width sweep & Vary compute and accumulation data types for the same operation & Outputs match within tolerance; deviations expose type promotion or casting bugs & RC5 \\ \hline

% \multirow{3}{2cm}{Invariant or Property Based Testing} 
%  & No-op IR transformations & Apply transformations such as common subexpression elimination, loop-invariant code motion, or dead code elimination that should preserve semantics & IR fingerprint and outputs must remain identical & RC2 \\ \cline{2-5}
%  & Algebraic identities & Verify monotonicity or symmetry of operators (e.g., $f(-x)=-f(x)$) & Deviations indicate incorrect operator implementation or fusion logic & RC5 \\ \cline{2-5}
%  & Determinism check & Repeated runs with identical inputs & Hash or checksum of outputs must match across runs & RC4 \\ \hline

 \multirow{3}{2cm}{Invariant or Property-Based Testing}
& No-op IR transformations & Apply semantics-preserving transformations  & IR structure and outputs must remain identical; divergence signals invalid rewrite & RC2 \\ \cline{2-5}
& Algebraic identity checks & Validate mathematical properties (e.g., $f(-x) = -f(x)$) & Violations reveal incorrect operator logic or fused rewrite errors & RC5 \\ \cline{2-5}
& Determinism check & Re-run kernel with same inputs and configurations & Output hashes must match; nondeterminism signals memory races or unguarded state & RC4 \\ \hline

% \multirow{3}{2cm}{Canary / Injection-based Testing} 
%  & Tail-region overwrite & Append fixed canary bytes beyond valid memory bounds & Canary region must remain unmodified; detects out-of-bounds writes & RC1, RC4 \\ \cline{2-5}
%  & Guard-region check & Surround buffers with sentinel values during execution & All sentinels preserved; violation flags memory overflow or aliasing & RC4 \\ \cline{2-5}
%  & Write-unique ID kernel & Write logical indices as outputs and verify bijection & Every logical index appears once; detects incorrect mapping logic & RC3 \\ \hline
 \multirow{3}{2cm}{Canary / Injection-Based Testing}
& Tail canary overwrite & Append fixed sentinel bytes after valid memory & Canaries must remain unchanged; detects out-of-bounds writes at tile edges & RC1, RC4 \\ \cline{2-5}
& Guard-region sentinels & Surround allocated buffers with known patterns & All sentinels preserved; flags memory overflow or illegal aliasing & RC4 \\ \cline{2-5}
& Unique ID writes & Emit logical indices and check output bijection & Every index must appear exactly once; reveals launch or mapping bugs & RC3 \\ \hline
\end{tabular}}
\label{tab:test-oracles}
\end{table*}

\section{Fixing Tile-Program Codegen Bugs}
\label{fixing}

We analyzed commit histories, developer comments, and bug reports to identify recurring repair strategies for each root-cause category. These strategies summarize observed developer practices rather than directly applicable repair procedures. Applying a repair strategy requires knowledge of the compiler's intermediate representations, transformation passes, and backend constraints. Table~\ref{tab:fix_totals} summarizes the identified strategies.

% \begin{table}[ht]
% \centering
% %\small
% %\setlength{\tabcolsep}{3pt}
% \caption{Summary of Fixing Strategies}
% \resizebox{\columnwidth}{!}{
% \begin{tabular}{|p{5cm}|c|c|}
% \hline
% \textbf{Fix Strategy} & \textbf{Triton Repo} & \textbf{Other Repos} \\
% \hline
% Change Index / Pointer Access   & 2 & 1 \\
% Fix Index Calculation           & 3 & -- \\
% Change Batch / Block Size       & 3 & 2  \\
% Add Architecture Support        & 2 & 3  \\
% Change Backend Implementation   & 13 & 10 \\
% Add Operator Checks             & 3 & 2  \\
% Fix Operator Implementation     & 3 & 1  \\
% Add Input Check                 & 1 & 2 \\
% Change Data Type / Add Support  & 5 & 4  \\
% Change Number of Warps          & 3 & -- \\
% \hline
% \textbf{Total}                  & \textbf{38} & \textbf{25}  \\
% \hline
% \end{tabular}
% \label{tab:fixes}
% }
% \end{table}

%\begin{document}
 
\begin{table}[htbp]
\centering
\caption{Summary of fixing strategies}
\label{tab:fix_totals}
\resizebox{0.5\textwidth}{!}{
\begin{tabular}{lr} % 'l' for left-aligned text, 'r' for right-aligned numbers
\toprule
\textbf{Fixing Strategy} & \textbf{Total} \\
\midrule
Branch and looping handling & 12 \\
Device backend specialization & 3 \\
Indexing and bounds repair & 27\\
Instruction ordering \& scheduling fixes & 7 \\
IR lowering and mapping repairs & 39\\
Kernel launch and tile tuning & 6 \\
Memory allocation and capacity repairs & 28 \\
Operator implementation and vectorization & 87 \\
Thread/block/tile handling & 22 \\
Type and numerical support & 70 \\
\bottomrule
\end{tabular}
}
\end{table}
 
%\end{document}

\textbf{Repairing Control-Flow and Scheduling Bugs:} Developers corrected loop bounds, branch predicates, synchronization points, and instruction schedules. These changes restored correct handling of tile boundaries and the required execution order across threads and warps.

\textbf{Repairing IR Construction and Transformation Bugs:}
Developers repaired IR generation, lowering, and rewrite passes. The repairs corrected invalid PTX or LLVM constructs and preserved type consistency, dominance relations, and predicate coverage.

\textbf{Repairing Tile-Mapping and Launch Bugs:}
Developers changed tile shapes, launch configurations, and thread-to-data mappings to restore complete iteration-space coverage, especially for partial tiles and prime-sized dimensions.

\textbf{Repairing Memory Bugs:}
Developers added bounds checks and corrected address calculations, allocation sizes, and alignment constraints. Some repairs changed buffer-reuse or key-generation logic to prevent unintended aliasing.

\textbf{Repairing Type and Operator Bugs:}
This category contained the largest number of repairs. Developers implemented missing intrinsics, corrected invalid intrinsic use, and repaired unstable floating-point expressions. Other changes corrected the handling of special values such as NaN and infinities, data-type conversions, rounding modes, and operator formulations.

\textbf{Repairing Device-Specific Bugs:}
Developers added architecture-specific lowering rules and device-capability checks. These changes enforced constraints on PTX versions, vector widths, register use, alignment, and shared-memory allocation across GPU generations.

Across all categories, most repairs modified codegen logic or transformation passes. In our corpus, memory and type repairs were generally localized, whereas control-flow and IR repairs more often spanned multiple compiler stages.

\section{Threats to Validity}
\label{threats}

%Our findings are derived from a manual analysis of 301 real-world codegen bug reports, filtered to ensure relevance, correctness, and fix confirmation. While our methodology included independent reviews and joint resolution of disagreements, several threats remain.

Our findings are based on a manual analysis of 301 publicly reported tile codegen bugs. We applied explicit inclusion criteria, included only reports with confirmed fixes, and used independent review followed by discussion among the authors. Despite these procedures, the study remains subject to threats concerning external validity, selection bias, construct validity, and scope.

%\textbf{External validity.} Tile programming is a relatively new framework, and its ecosystem is still maturing. The number of publicly available repositories and bug reports is limited compared to established GPU ecosystems like CUDA or DL frameworks. As a result, our dataset may not fully represent the complete spectrum of bugs that will emerge as adoption grows. Additionally, the lack of prior studies on tile-specific codegen bugs constrains comparative generalizability.

\textbf{External validity.} Tile programming frameworks constitute a comparatively young and evolving ecosystem. Public repositories and confirmed bug reports are fewer than those available for mature GPU ecosystems such as CUDA and major DL frameworks. Our corpus may therefore omit bug patterns that occur in less visible projects or emerge as adoption expands. The absence of prior empirical studies of tile codegen bugs also prevents a direct comparison between our taxonomy and an independent corpus.

%\textbf{Selection bias}. Our bug corpus was collected from official and community-tracked tile programming projects, relying on publicly available issues and confirmed fixes. This excludes silent failures or unreported bugs in private codebases. While we enforced strict criteria for bug validity (e.g., closed status and linked patches), some edge cases may have been missed due to incomplete descriptions or ambiguous metadata.

\textbf{Selection bias.} We collected bug reports from official and community-maintained repositories and included only publicly documented reports that satisfied our relevance criteria and contained a confirmed fix. As a result, the corpus excludes unreported failures, including failures in private codebases and silent failures that developers did not detect or report. The selection criteria may also exclude valid reports with incomplete or ambiguous descriptions, patches, or metadata.

%\textbf{Construct validity.} Root cause classification was guided by prior work on CUDA and DL bug taxonomies but adapted to tile frameworks' unique programming model. Though reviewed by multiple authors, categorization remains subject to interpretation, especially in cases where abstraction layers obscure the exact failure mechanism.

%\textbf{Construct validity.} We adapted prior CUDA and DL bug taxonomies to the tile programming model and classified each report using its issue description, discussion, and associated patch. Multiple authors independently reviewed the classifications and discussed disagreements until reaching agreement. However, root-cause classification still requires interpretation, particularly when framework abstractions obscure the compiler stage or transformation responsible for a failure. Some bugs may therefore support more than one plausible classification.

\textbf{Construct validity.} We adapted prior CUDA and DL bug taxonomies to the tile programming frameworks and classified each report using its issue description, discussion, and patch. Multiple authors independently reviewed the classifications and discussed disagreements. However, root-cause classification remains interpretive, particularly when framework abstractions obscure the compiler stage or transformation responsible for a failure.

%\textbf{Scope limitations}. Our goal was not to exhaustively characterize all tile codegen bugs but to surface representative patterns of bugs, their causes, and fixes. Findings are qualitative and exploratory, offering a foundation for future tool development and empirical studies.

\textbf{Scope limitations.} Our study characterizes recurring root causes, symptoms, triggering conditions, and fixes in the collected corpus. The reported frequencies apply only to this corpus and do not estimate prevalence across all tile frameworks or deployed systems. The qualitative findings provide an empirical basis for developing future testing and debugging tools for tile codegen.

\section{Related Work}
\label{related}

%Prior research has extensively investigated bugs in GPU programs, particularly in handwritten CUDA kernels and low-level compiler backends~\cite{wu2020simulee, barracuda, gklee, gpuverify, boyer2008automated, islam2018bugaroo, sorensen2016exposing,jiang2020cudasmith, binfpe, laguna2019fpchecker, gpu-fpx,laguna2022finding, rathnasuriya2025investigation}. These studies primarily focus on developer-introduced memory errors, synchronization faults, and numerical inconsistencies. Tools such as GPUVerify~\cite{gpuverify} and GKLEE~\cite{gklee} perform static verification, while CUDA-memcheck and racecheck instrument runtime execution to detect out-of-bounds accesses and shared-memory races~\cite{gerfin2012debugging}. Simulee and GRace~\cite{wu2020simulee} combine guided input generation with dynamic analysis to uncover concurrency issues, and CUDAsmith~\cite{jiang2020cudasmith} applies differential testing to detect backend miscompilations. Separate efforts have explored numerical correctness in GPU code~\cite{binfpe,laguna2019fpchecker, laguna2022finding}, including floating-point errors introduced by limited precision and backend-specific transformations.

We organize related research into four areas: GPU program analysis, compiler testing, DL framework testing, and tile-based systems. Prior research on GPU program bugs primarily examines handwritten CUDA kernels~\cite{wu2020simulee, barracuda, gklee, gpuverify, boyer2008automated, islam2018bugaroo, sorensen2016exposing, binfpe, laguna2019fpchecker, gpu-fpx, laguna2022finding, rathnasuriya2025investigation}. These studies investigate developer-introduced memory errors, synchronization faults, and numerical inconsistencies. GPUVerify~\cite{gpuverify} and GKLEE~\cite{gklee} statically verify memory and synchronization properties. Runtime tools such as CUDA-memcheck and Racecheck instrument program execution to detect out-of-bounds memory accesses and shared-memory races~\cite{gerfin2012debugging}. Simulee and GRace~\cite{wu2020simulee} combine guided input generation with dynamic analysis to expose concurrency bugs. Other techniques detect floating-point errors caused by limited precision or backend-specific transformations~\cite{binfpe, laguna2019fpchecker, gpu-fpx, laguna2022finding}. These techniques target bugs in handwritten kernels or their execution rather than bugs introduced when a tile compiler constructs control flow, schedules, predicates, and memory layouts.

GPU compiler testing detects miscompilations through random program generation, metamorphic testing, and differential testing. Representative techniques include CUDAsmith for CUDA~\cite{jiang2020cudasmith}, CLsmith for OpenCL~\cite{lidbury2015clsmith}, GLFuzz~\cite{donaldson2017glfuzz} and ShaDiv~\cite{shadiv} for the OpenGL Shading Language (GLSL), and DarthShader~\cite{darthshader} and WGSLsmith~\cite{wgslsmith} for WebGPU. Tile compilers may emit lower-level representations such as PTX or SPIR-V, which downstream GPU compiler toolchains then process. GPU compiler fuzzers may therefore expose a tile codegen bug after the bug propagates to a downstream stage. However, these fuzzers do not systematically control tile-level inputs such as tile sizes, memory layouts, scheduling parameters, or shape-dependent lowering decisions. A downstream failure also does not identify the tile-level compilation decision that introduced the bug.

Studies of DL frameworks such as TensorFlow and PyTorch examine operator-level issues, including incorrect tensor shapes, parameter mismatches, and inconsistencies across frameworks or backend configurations~\cite{DLGPU1, DLGPU2, DLGPU3, DLGPU4, DLGPU6, DLGPU7, DLGPUtenfuzz, DLGPUaudee}. TenFuzz~\cite{DLGPUtenfuzz} mutates unit tests to trigger shape and type errors. CRADLE~\cite{DLGPU4} and Audee~\cite{DLGPUaudee} perform differential testing across frameworks or backend configurations. This line of research focuses on operator integration and cross-framework or cross-backend inconsistencies.

DL framework tests can reach tile compilation through operator dispatch or model compilation, particularly because DL backends increasingly use Triton-based compilation and tile-based operator libraries. However, these techniques do not systematically control tile-level parameters such as tile shapes, mask coverage at partial-tile boundaries, warp counts, or shared-memory staging. When a DL-level test triggers a tile codegen bug, the failure typically appears as an operator-level output discrepancy or backend inconsistency. Such a test does not localize the failure to the tile-generation decision that introduced it.

%Compiler-focused research in LLVM, MLIR, and GCC investigates miscompilations, undefined behavior, and transformation bugs~\cite{yang2011finding,livinskii2020random,ma2023survey,even2023grayc,limpanukorn2024fuzzing,wang2025duoreduce,wang2023mlirsmith,suo2025desil}. Csmith stress tests backend correctness using random program generation, while MLIRSmith and DESIL examine lowering path inconsistencies and undefined semantics in structured IRs. However, these tools are designed for imperative control-flow programs and lack mechanisms for tiling, predication, or hardware-specific indexing.

Compiler-focused research on LLVM, MLIR, and GCC investigates miscompilations, undefined behavior, and transformation bugs~\cite{yang2011finding, livinskii2020random, ma2023survey, even2023grayc, limpanukorn2024fuzzing, wang2025duoreduce, wang2023mlirsmith, suo2025desil}. Csmith tests backend correctness through random program generation, whereas MLIRSmith and DESIL examine lowering-path inconsistencies and undefined semantics in structured intermediate representations. However, these tools primarily target imperative control-flow programs and do not provide mechanisms for systematically varying tiling, predication, or hardware-specific indexing decisions.

%Existing methodologies are not directly applicable to tile-based codegen. Tile programming frameworks rely on compiler-driven synthesis of control flow, tiling, and memory layout, introducing transformation-induced bugs absent in handwritten kernels. Prior work focuses on performance optimization and backend fusion~\cite{beck2025tiled,zheng2025tilelink,yuan2025native,gupta2025splat,zhang2025flash} rather than bug diagnosis or symptom analysis.

GPU-program analyzers, DL-framework testing techniques, and general compiler fuzzers therefore do not directly address tile-based codegen. Tile programming frameworks rely on the compiler to construct control flow, select tiling strategies, and determine memory layouts. Errors in these compiler-generated decisions produce transformation-induced bugs that differ from developer-introduced bugs in handwritten kernels. Research on tile-based systems primarily investigates performance optimization and backend fusion~\cite{beck2025tiled, zheng2025tilelink, yuan2025native, gupta2025splat, zhang2025flash}, rather than bug diagnosis or symptom analysis. Existing benchmarks also evaluate the correctness and performance of GPU kernels generated by large language models~\cite{li2025tritonbench, ouyang2025kernelbench}, but they do not examine bugs introduced within tile compilation pipelines.

%To our knowledge, there has been no systematic study of real-world bugs in tile-based codegen. Existing work primarily focuses on large language model–based approaches to generate kernel code but does not address the correctness and reliability of tile-based compilation pipelines~\cite{li2025tritonbench,ouyang2025kernelbench}. Our work fills this gap by analyzing 301 bug reports, constructing a root-cause taxonomy, and identifying how tile-specific abstractions influence bug manifestation, detection, and repair. We further observe that no existing testing tools are designed for tile codegen, leaving a critical gap in compiler validation. The findings in this paper would serve as a foundation for future community-driven development of dedicated testing and debugging infrastructure for tile compiler pipelines. 

To the best of our knowledge, no prior study systematically examines real-world tile codegen bugs. Our study addresses this gap by analyzing 301 tile codegen bugs, constructing a root-cause taxonomy, and characterizing how these bugs manifest, what triggers them, and how developers detect, diagnose, and fix them. These findings provide an empirical basis for testing and debugging techniques designed specifically for tile codegen.

\section{Conclusion}
\label{conclusion}

In this work, we present the first empirical study of codegen bugs in tile-based programming frameworks, uncovering their root causes, manifestation patterns, and detection challenges. Our analysis spans 301 real-world bugs and reveals that many bugs arise from the interaction between high-level abstractions and low-level hardware constraints, leading to subtle bugs in control-flow, memory, tiling, and type handling. To facilitate detection, we summarize bug detection strategies leveraging test input generation and test oracles.  We believe this study lays a foundation for future research on verification, testing, and tooling for next-generation accelerator compilers.

\section*{Data Availability}
\label{data}
The artifact repository ~\cite{tritonbug2025repo} contains the datasets and code needed to reproduce the results in this paper. Our project website is available at ~\cite{tritonbug2025website}.

% \section*{ACKNOWLEDGMENTS}
% This work was partially supported by NSF grants NSF CCF-2146443 and Amazon Trust AI Research Award. We extend our gratitude to Mirazul Haque for hiss guidance throughout this project.

%%%oopsla
%We confirm that the source code of the experiments is available at an anonymous GitHub repository for peer review. The GitHub repository includes all source codes to run the experiments and scripts to download the datasets and models for producing the experiment results. The GitHub repository is available at: \href{https://github.com/CodeImprove/CodeImprove/tree/main}{https://github.com/CodeImprove/CodeImprove/tree/main}

%\input{tex/dataavailability}

%% Acknowledgments
\begin{acks}                            %% acks environment is optional
                                        %% contents suppressed with 'anonymous'
  %% Commands \grantsponsor{<sponsorID>}{<name>}{<url>} and
  %% \grantnum[<url>]{<sponsorID>}{<number>} should be used to
  %% acknowledge financial support and will be used by metadata
  %% extraction tools.
  % This material is based upon work supported by the
  % \grantsponsor{GS100000001}{National Science
  %   Foundation}{http://dx.doi.org/10.13039/100000001} under Grant
  % No.~\grantnum{GS100000001}{nnnnnnn} and Grant
  % No.~\grantnum{GS100000001}{mmmmmmm}.  Any opinions, findings, and
  % conclusions or recommendations expressed in this material are those
  % of the author and do not necessarily reflect the views of the
  % National Science Foundation.
  This work was supported in part by the Fundamental and Interdisciplinary Disciplines Breakthrough Plan of the Ministry of Education of China (No. JYB2025XDXM118), the National Natural Science Foundation of China under Grant Nos. U25A6023 and 92464301, and an Amazon Trust AI Research Award. Any opinions, findings, and conclusions or recommendations expressed in this material are those of the authors and do not necessarily reflect the views of the funding agencies.
\end{acks}

%% Bibliography
\bibliographystyle{ACM-Reference-Format}
\bibliography{main}

%% Appendix
%\appendix
%\section{Appendix}

%Text of appendix \ldots

\end{document}